\documentclass[twocolumn]{aastex63}

\usepackage{amsmath}
\usepackage{amssymb}
\usepackage{amsthm}
\usepackage{booktabs}
\usepackage{graphicx}
\usepackage[colorinlistoftodos]{todonotes}
\usepackage{natbib}

\usepackage{lineno}
\usepackage{subfigure}
\usepackage{soul}
\usepackage{verbatim}

\setwatermarkfontsize{0.75in}
\definecolor{darkgreen}{rgb}{0.05,0.3,0.05}

\begin{document}

\vspace*{-\headsep}\vspace*{\headheight}
{\footnotesize \hfill FERMILAB-PUB-22-243-SCD}\\
\vspace*{-\headsep}\vspace*{\headheight}
{\footnotesize \hfill DES-2021-0682} 

\title{DeepZipper II: Searching for Lensed Supernovae in Dark Energy Survey Data with Deep Learning}

\correspondingauthor{Robert Morgan}
\email{robert.morgan@wisc.edu}

\author[0000-0002-7016-5471]{R.~Morgan}
\affil{Physics Department, University of Wisconsin-Madison, Madison, WI  53706, USA}
\affil{Fermi National Accelerator Laboratory, P. O. Box 500, Batavia, IL 60510, USA}
\affil{Legacy Survey of Space and Time Corporation Data Science Fellowship Program, USA}

\author[0000-0001-6706-8972]{B.~Nord}
\affil{Fermi National Accelerator Laboratory, P. O. Box 500, Batavia, IL 60510, USA}
\affil{Kavli Institute for Cosmological Physics, University of Chicago, Chicago, IL 60637, USA}

\author[0000-0001-8156-0429]{K.~Bechtol}
\affil{Physics Department, University of Wisconsin-Madison, Madison, WI  53706, USA}
\affil{Legacy Survey of Space and Time, 933 North Cherry Avenue, Tucson, AZ 85721, USA}

\author[0000-0001-8211-8608]{A.~M\"oller}
\affil{Centre for Astrophysics \& Supercomputing, Swinburne University of Technology, Victoria 3122, Australia}

\author{W.~G.~Hartley}
\affil{Department of Astronomy, University of Geneva, ch. d’Ecogia 16, CH-1290 Versoix, Switzerland}

\author[0000-0003-3195-5507]{S.~Birrer}
\affil{Kavli Institute for Particle Astrophysics and Cosmology, Department of Physics, Stanford University, Stanford, CA 94305, USA}
\affil{SLAC National Accelerator Laboratory, Menlo Park, CA 94025, USA}

\author[0000-0001-7282-3864]{S.~J.~Gonz\'{a}lez}
\affil{Physics Department, University of Wisconsin-Madison, Madison, WI  53706, USA}

\author[0000-0002-8397-8412]{M.~Martinez}
\affil{Physics Department, University of Wisconsin-Madison, Madison, WI  53706, USA}

\author{R.~A.~Gruendl}
\affil{Center for Astrophysical Surveys, National Center for Supercomputing Applications, 1205 West Clark St., Urbana, IL 61801, USA}
\affil{Department of Astronomy, University of Illinois at Urbana-Champaign, 1002 W. Green Street, Urbana, IL 61801, USA}

\author[0000-0002-3304-0733]{E.~J.~Buckley-Geer}
\affil{Fermi National Accelerator Laboratory, P. O. Box 500, Batavia, IL 60510, USA}
\affil{Department of Astronomy and Astrophysics, University of Chicago, Chicago, IL 60637, USA}

\author{A.~J.~Shajib}
\affil{Department of Astronomy and Astrophysics, University of Chicago, Chicago, IL 60637, USA}
\affil{NHFP Einstein Fellow}

\author[0000-0003-3044-5150]{A.~Carnero~Rosell}
\affil{Instituto de Astrofisica de Canarias, E-38205 La Laguna, Tenerife, Spain}
\affil{Laborat\'orio Interinstitucional de e-Astronomia - LIneA, Rua Gal. Jos\'e Cristino 77, Rio de Janeiro, RJ - 20921-400, Brazil}
\affil{Universidad de La Laguna, Dpto. Astrofísica, E-38206 La Laguna, Tenerife, Spain}

\author[0000-0003-1731-0497]{C.~Lidman}
\affil{Centre for Gravitational Astrophysics, College of Science, The Australian National University, ACT 2601, Australia}
\affil{The Research School of Astronomy and Astrophysics, The Australian National University, ACT 2601, Australia}

\author{T.~Collett}
\affil{Institute of Cosmology and Gravitation, University of Portsmouth, Portsmouth, PO1 3FX, UK}


\author{T.~M.~C.~Abbott}
\affil{Cerro Tololo Inter-American Observatory, NSF's National Optical-Infrared Astronomy Research Laboratory, Casilla 603, La Serena, Chile}

\author{M.~Aguena}
\affil{Laborat\'orio Interinstitucional de e-Astronomia - LIneA, Rua Gal. Jos\'e Cristino 77, Rio de Janeiro, RJ - 20921-400, Brazil}

\author{F.~Andrade-Oliveira}
\affil{Department of Physics, University of Michigan, Ann Arbor, MI 48109, USA}

\author[0000-0002-0609-3987]{J.~Annis}
\affil{Fermi National Accelerator Laboratory, P. O. Box 500, Batavia, IL 60510, USA}

\author{D.~Bacon}
\affil{Institute of Cosmology and Gravitation, University of Portsmouth, Portsmouth, PO1 3FX, UK}

\author[0000-0002-4900-805X]{S.~Bocquet}
\affil{University Observatory, Faculty of Physics, Ludwig-Maximilians-Universit\"at, Scheinerstr. 1, 81679 Munich, Germany}

\author[0000-0002-8458-5047]{D.~Brooks}
\affil{Department of Physics \& Astronomy, University College London, Gower Street, London, WC1E 6BT, UK}

\author{D.~L.~Burke}
\affil{Kavli Institute for Particle Astrophysics \& Cosmology, P. O. Box 2450, Stanford University, Stanford, CA 94305, USA}
\affil{SLAC National Accelerator Laboratory, Menlo Park, CA 94025, USA}

\author[0000-0002-4802-3194]{M.~Carrasco~Kind}
\affil{Center for Astrophysical Surveys, National Center for Supercomputing Applications, 1205 West Clark St., Urbana, IL 61801, USA}
\affil{Department of Astronomy, University of Illinois at Urbana-Champaign, 1002 W. Green Street, Urbana, IL 61801, USA}

\author[0000-0002-3130-0204]{J.~Carretero}
\affil{Institut de F\'{\i}sica d'Altes Energies (IFAE), The Barcelona Institute of Science and Technology, Campus UAB, 08193 Bellaterra (Barcelona) Spain}

\author[0000-0001-7316-4573]{F.~J.~Castander}
\affil{Institut d'Estudis Espacials de Catalunya (IEEC), 08034 Barcelona, Spain}
\affil{Institute of Space Sciences (ICE, CSIC),  Campus UAB, Carrer de Can Magrans, s/n,  08193 Barcelona, Spain}

\author[0000-0003-1949-7638]{C.~Conselice}
\affil{Jodrell Bank Center for Astrophysics, School of Physics and Astronomy, University of Manchester, Oxford Road, Manchester, M13 9PL, UK}
\affil{University of Nottingham, School of Physics and Astronomy, Nottingham NG7 2RD, UK}

\author{L.~N.~da Costa}
\affil{Laborat\'orio Interinstitucional de e-Astronomia - LIneA, Rua Gal. Jos\'e Cristino 77, Rio de Janeiro, RJ - 20921-400, Brazil}
\affil{Observat\'orio Nacional, Rua Gal. Jos\'e Cristino 77, Rio de Janeiro, RJ - 20921-400, Brazil}

\author{M.~Costanzi}
\affil{Astronomy Unit, Department of Physics, University of Trieste, via Tiepolo 11, I-34131 Trieste, Italy}
\affil{INAF-Osservatorio Astronomico di Trieste, via G. B. Tiepolo 11, I-34143 Trieste, Italy}
\affil{Institute for Fundamental Physics of the Universe, Via Beirut 2, 34014 Trieste, Italy}

\author[0000-0001-8318-6813]{J.~De~Vicente}
\affil{Centro de Investigaciones Energ\'eticas, Medioambientales y Tecnol\'ogicas (CIEMAT), Madrid, Spain}

\author[0000-0002-0466-3288]{S.~Desai}
\affil{Department of Physics, IIT Hyderabad, Kandi, Telangana 502285, India}

\author{P.~Doel}
\affil{Department of Physics \& Astronomy, University College London, Gower Street, London, WC1E 6BT, UK}

\author{S.~Everett}
\affil{Santa Cruz Institute for Particle Physics, Santa Cruz, CA 95064, USA}

\author{I.~Ferrero}
\affil{Institute of Theoretical Astrophysics, University of Oslo. P.O. Box 1029 Blindern, NO-0315 Oslo, Norway}

\author[0000-0002-2367-5049]{B.~Flaugher}
\affil{Fermi National Accelerator Laboratory, P. O. Box 500, Batavia, IL 60510, USA}

\author{D.~Friedel}
\affil{Center for Astrophysical Surveys, National Center for Supercomputing Applications, 1205 West Clark St., Urbana, IL 61801, USA}

\author[0000-0003-4079-3263]{J.~Frieman}
\affil{Fermi National Accelerator Laboratory, P. O. Box 500, Batavia, IL 60510, USA}
\affil{Kavli Institute for Cosmological Physics, University of Chicago, Chicago, IL 60637, USA}

\author[0000-0002-9370-8360]{J.~Garc\'ia-Bellido}
\affil{Instituto de Fisica Teorica UAM/CSIC, Universidad Autonoma de Madrid, 28049 Madrid, Spain}

\author[0000-0001-9632-0815]{E.~Gaztanaga}
\affil{Institut d'Estudis Espacials de Catalunya (IEEC), 08034 Barcelona, Spain}
\affil{Institute of Space Sciences (ICE, CSIC),  Campus UAB, Carrer de Can Magrans, s/n,  08193 Barcelona, Spain}

\author[0000-0003-3270-7644]{D.~Gruen}
\affil{University Observatory, Faculty of Physics, Ludwig-Maximilians-Universit\"at, Scheinerstr. 1, 81679 Munich, Germany}

\author[0000-0003-0825-0517]{G.~Gutierrez}
\affil{Fermi National Accelerator Laboratory, P. O. Box 500, Batavia, IL 60510, USA}

\author{S.~R.~Hinton}
\affil{School of Mathematics and Physics, University of Queensland, Brisbane, QLD 4072, Australia}

\author{D.~L.~Hollowood}
\affil{Santa Cruz Institute for Particle Physics, Santa Cruz, CA 95064, USA}

\author[0000-0002-6550-2023]{K.~Honscheid}
\affil{Center for Cosmology and Astro-Particle Physics, The Ohio State University, Columbus, OH 43210, USA}
\affil{Department of Physics, The Ohio State University, Columbus, OH 43210, USA}

\author[0000-0003-0120-0808]{K.~Kuehn}
\affil{Australian Astronomical Optics, Macquarie University, North Ryde, NSW 2113, Australia}
\affil{Lowell Observatory, 1400 Mars Hill Rd, Flagstaff, AZ 86001, USA}

\author[0000-0003-2511-0946]{N.~Kuropatkin}
\affil{Fermi National Accelerator Laboratory, P. O. Box 500, Batavia, IL 60510, USA}

\author[0000-0002-1134-9035]{O.~Lahav}
\affil{Department of Physics \& Astronomy, University College London, Gower Street, London, WC1E 6BT, UK}

\author{M.~Lima}
\affil{Departamento de F\'isica Matem\'atica, Instituto de F\'isica, Universidade de S\~ao Paulo, CP 66318, S\~ao Paulo, SP, 05314-970, Brazil}
\affil{Laborat\'orio Interinstitucional de e-Astronomia - LIneA, Rua Gal. Jos\'e Cristino 77, Rio de Janeiro, RJ - 20921-400, Brazil}

\author[0000-0002-1372-2534]{F.~Menanteau}
\affil{Center for Astrophysical Surveys, National Center for Supercomputing Applications, 1205 West Clark St., Urbana, IL 61801, USA}
\affil{Department of Astronomy, University of Illinois at Urbana-Champaign, 1002 W. Green Street, Urbana, IL 61801, USA}

\author[0000-0002-6610-4836]{R.~Miquel}
\affil{Instituci\'o Catalana de Recerca i Estudis Avan\c{c}ats, E-08010 Barcelona, Spain}
\affil{Institut de F\'{\i}sica d'Altes Energies (IFAE), The Barcelona Institute of Science and Technology, Campus UAB, 08193 Bellaterra (Barcelona) Spain}

\author[0000-0002-6011-0530]{A.~Palmese}
\affil{Department of Astronomy, University of California, Berkeley,  501 Campbell Hall, Berkeley, CA 94720, USA}

\author{F.~Paz-Chinch\'{o}n}
\affil{Center for Astrophysical Surveys, National Center for Supercomputing Applications, 1205 West Clark St., Urbana, IL 61801, USA}
\affil{Institute of Astronomy, University of Cambridge, Madingley Road, Cambridge CB3 0HA, UK}

\author{M.~E.~S.~Pereira}
\affil{Hamburger Sternwarte, Universit\"{a}t Hamburg, Gojenbergsweg 112, 21029 Hamburg, Germany}

\author[0000-0001-9186-6042]{A.~Pieres}
\affil{Laborat\'orio Interinstitucional de e-Astronomia - LIneA, Rua Gal. Jos\'e Cristino 77, Rio de Janeiro, RJ - 20921-400, Brazil}
\affil{Observat\'orio Nacional, Rua Gal. Jos\'e Cristino 77, Rio de Janeiro, RJ - 20921-400, Brazil}

\author[0000-0002-2598-0514]{A.~A.~Plazas~Malag\'on}
\affil{Department of Astrophysical Sciences, Princeton University, Peyton Hall, Princeton, NJ 08544, USA}

\author{J.~Prat}
\affil{Department of Astronomy and Astrophysics, University of Chicago, Chicago, IL 60637, USA}
\affil{Kavli Institute for Cosmological Physics, University of Chicago, Chicago, IL 60637, USA}

\author{M.~Rodriguez-Monroy}
\affil{Centro de Investigaciones Energ\'{e}ticas, Medioambientales y Tecnol\'{o}gicas (CIEMAT), Madrid, Spain}

\author[0000-0002-9328-879X]{A.~K.~Romer}
\affil{Department of Physics and Astronomy, Pevensey Building, University of Sussex, Brighton, BN1 9QH, UK}

\author[0000-0001-5326-3486]{A.~Roodman}
\affil{Kavli Institute for Particle Astrophysics \& Cosmology, P. O. Box 2450, Stanford University, Stanford, CA 94305, USA}
\affil{SLAC National Accelerator Laboratory, Menlo Park, CA 94025, USA}

\author[0000-0002-9646-8198]{E.~Sanchez}
\affil{Centro de Investigaciones Energ\'eticas, Medioambientales y Tecnol\'ogicas (CIEMAT), Madrid, Spain}

\author{V.~Scarpine}
\affil{Fermi National Accelerator Laboratory, P. O. Box 500, Batavia, IL 60510, USA}

\author[0000-0002-1831-1953]{I.~Sevilla-Noarbe}
\affil{Centro de Investigaciones Energ\'eticas, Medioambientales y Tecnol\'ogicas (CIEMAT), Madrid, Spain}

\author[0000-0002-3321-1432]{M.~Smith}
\affil{School of Physics and Astronomy, University of Southampton,  Southampton, SO17 1BJ, UK}

\author[0000-0002-7047-9358]{E.~Suchyta}
\affil{Computer Science and Mathematics Division, Oak Ridge National Laboratory, Oak Ridge, TN 37831}

\author{M.~E.~C.~Swanson}
\affil{School of Physics and Astronomy, University of Southampton, Southampton, SO17 1BJ, UK}

\author[0000-0003-1704-0781]{G.~Tarle}
\affil{Department of Physics, University of Michigan, Ann Arbor, MI 48109, USA}

\author{D.~Thomas}
\affil{Institute of Cosmology and Gravitation, University of Portsmouth, Portsmouth, PO1 3FX, UK}

\author{T.~N.~Varga}
\affil{Excellence Cluster Origins, Boltzmannstr.\ 2, 85748 Garching, Germany}
\affil{Max Planck Institute for Extraterrestrial Physics, Giessenbachstrasse, 85748 Garching, Germany}
\affil{Universit\"ats-Sternwarte, Fakult\"at f\"ur Physik, Ludwig-Maximilians Universit\"at M\"unchen, Scheinerstr. 1, 81679 M\"unchen, Germany}

\begin{abstract}
Gravitationally lensed supernovae (LSNe) are important probes of cosmic expansion, but they remain rare and difficult to find.
Current cosmic surveys likely contain 5-10 LSNe in total while next-generation experiments are expected to contain several hundreds to a few thousands of these systems.
We search for these systems in observed Dark Energy Survey (DES) 5-year SN fields -- 10 3-sq.~deg. regions of sky imaged in the $griz$ bands approximately every six nights over five years.
To perform the search, we utilize the DeepZipper approach: a multi-branch deep learning architecture trained on image-level simulations of LSNe that simultaneously learns spatial and temporal relationships from time series of images.
We find that our method obtains a LSN recall of 61.13\% and a false positive rate of 0.02\% on the DES SN field data.
DeepZipper selected 2,245 candidates from a magnitude-limited ($m_i$ $<$ 22.5) catalog of 3,459,186 systems.
We employ human visual inspection to review systems selected by the network and find three candidate LSNe in the DES SN fields.
\end{abstract}

\keywords{Optical astronomy -- Machine learning -- Transient sources -- Strong lensing}


\section{Introduction}
\label{sec:introduction}

Galaxy-scale gravitational lensing occurs when the gravitational potential of a foreground galaxy (positioned along an observer's line of sight to a background galaxy) is large enough to deflect the photons of the background galaxy on their journey to an observer.
This process produces arcs and/or multiple images of the background galaxy \citep{slreview}.
For the specific case in which the background galaxy contains a supernova (SN), the photons that contribute to each of the multiple images of the lensed supernova (LSN) travel different paths and distances to the observer and encounter different depths of gravitational potential depending on the distribution of the foreground galaxy's mass.
Because the speed of light is constant, the distinct paths correspond to distinct arrival times of the photons from each SN image.
Combining this time delay with a model of the foreground galaxy's mass distribution enables the direct inference of the rate of expansion of the Universe today $H_0$, as well as other cosmological parameters \citep{tdcosmography}.

Historically, LSNe are rare --- only a few detections have been made in total \citep{kelly, Rodney2021, iptf16geu, unresolved_lsn_1, unresolved_lsn_2, unresolved_lsn_3}.
However, modern optical time-domain survey datasets, such as those collected in the southern hemisphere by the Dark Energy Survey's SN fields \citep[DES;][]{desoverview, desfinale}, in the northern hemisphere by the Zwicky Transient Facility \citep{ztf} and the Young Supernova Experiment \citep{yse}, and over the next decade by the Vera C. Rubin Observatory's Legacy Survey of Space and Time \citep[LSST;][]{lsst}, are promising places to search for LSNe.
Based on imaging depth, sky area, and duration of observations, the DES SN fields are expected to contain $\sim0.5-2$ LSNe, and the LSST wide field is expected to contain $\sim2,000$ LSNe \citep{oguri}.
These datasets, which contain hundreds of millions to tens of billions of objects that are not LSNe, pose a significant challenge for searches \citep{desdr2, lsst_table}. 
In particular, it is vital to identify an LSN rapidly to enable follow-up observations before the SN fades during the weeks to months after the explosion \citep{sn_decay}. 
To keep pace with the data streams of large surveys and identify candidate LSNe promptly, we require fast and robust algorithms.

In \citet[]{deepzipper} -- hereafter referred to as ``DZ1'' -- we designed a deep learning detection architecture (``ZipperNet'') for LSNe  and demonstrated its performance on four simulated optical survey datasets that mimic DES and LSST.
In this work, we use a ZipperNet to search the DES SN fields \citep{desdr2} for LSNe.
We also discuss the data collection and data reduction steps necessary to carry out a comprehensive LSN search in an optical survey dataset.
We have made all code for data processing and deep learning available at \citep{zenodo_dz2}.

We present this work as follows. 
In Section \ref{sec:data}, we describe the characteristics of the DES SN field data. 
In Section \ref{sec:deeplearning}, we describe the training and optimization of our deep learning approach.
In Section \ref{sec:results}, we quantify the performance of this architecture on the DES SN field data, as well as present candidate LSN systems.
In Section \ref{sec:discussion}, we discuss the significance of the results and the outlook for detecting LSNe in Rubin Observatory data.
We conclude in Section \ref{sec:conclusion}.

\section{Data Collection}
\label{sec:data}

\subsection{The DES SN Fields}
\label{sec:deepfields}

DES SN field data were collected a) to facilitate the Type Ia-SN (SN-Ia) cosmology analyses in DES that use the single-epoch images and b) to enable galaxy population modeling (near the detection limits of the DES wide-field survey) that uses coadded images.
All data were collected with DECam \citep{brenna} on the Victor M. Blanco telescope from the Cerro-Tololo Inter-American Observatory in Chile between 2012 and 2018.
There are 10 3-sq.~deg. fields -- eight \textit{shallow} fields (X1, X2, E1, E2, C1, C2, S1, and S2) observed to a single-visit depth of $\sim23.5$~mag and two \textit{deep} fields (X3 and C3) observed to a single-visit depth of $\sim24.5$~mag.
Each field was imaged in the $griz$ bands approximately every six nights over five years, subject to Sun, Moon, and weather conditions.
The median full-width-at-half-maximum point spread functions (``seeing'') for the SN field images used in this analysis (after the downsampling discussed in section \ref{sec:objectselection}) were 1.37\arcsec, 1.26\arcsec, 1.15\arcsec, and 1.08\arcsec\ for the $griz$ bands, respectively.

\subsection{Candidate System Selection and Data Reduction}
\label{sec:objectselection}

We begin our search for candidate LSNe with all cataloged objects of DES Data Release 1 (also referred to as the ``Year 3 Gold Catalog'') \citep{desdr1}.
We construct an initial sample by requiring the object to be positioned within one of the SN fields and requiring all $griz$ \texttt{MAG\_AUTO} measurements to be brighter than 27.5~mag.
Then, within that sample, we require the $i$ band \texttt{MAG\_AUTO} only to be brighter than 22.5~mag to restrict the total number of objects in this first search of the DES SN fields.
Also within the initial sample, we require a catalog-level parameter size measurement (\texttt{CM\_T}) to be greater than 0.05, which excludes non-extended objects (e.g., stars) with approximately 99\% galaxy purity and 98\% galaxy completeness.
To evaluate the purity and completeness, we take a nearest-neighbor machine learning classifier that combines DES photometry with near-infrared photometry as truth, which has shown near-perfect performance at mag$_i < 22.5$ \citep{desdeep}.
These cuts produce a sample of 3,459,186 candidate systems for our analysis.

We next introduce a selection on the images that are used in the LSNe search across all five years of DES SN field exposures \citep{desdr2}.
If a system has two images on the same night in the same band, we choose the image for which the object was observed with the better seeing.
For each image, we also require the cataloged object's centroid to be positioned more than 23 pixels from all CCD edges: this permits constructing image cutouts (45 pixel by 45 pixel) without producing partial images.
Finally, to enforce cadence uniformity and simplify data processing, we require the same number of observations in each of the $griz$ bands.
We determine the band with the fewest useful observations and exclude images from the other bands to match it.
In doing so, we exclude images from regions of the time series in descending order of the sampling rate.
Thus, for each candidate lens galaxy in the SN fields selected from the DES catalog, we obtain a time series image set with the same number of images in each band of $griz$.
A typical length for a time series image set is $\sim20-35$ epochs.
We process each year of DES data independently.

\section{Deep Learning Methods}
\label{sec:deeplearning}

\subsection{Training Set Construction}

Our approach for detecting LSNe in the DES deep fields requires samples of LSNe (positives) and non-LSNe (negatives) to train the ZipperNet in a binary classification scheme.
To construct the training set, we used $\sim2$\% of the total dataset -- 76,203 time series image sets.
Due to the lack of real LSN examples, we create the positive class by using gravitational lensing simulation software \citep[\texttt{deeplenstronomy};][]{deeplenstronomy} to add LSNe to DES images in the training set.
For the negative class, we use time series image sets selected at random from the dataset.
Even given the erroneous case where a real LSN is randomly selected for the negative training class, LSNe are expected to be sufficiently rare in the DES SN fields such that this error would be infrequent and not affect the training.
Nevertheless, the two most likely types of false positives will be non-lensed SNe and strongly lensed galaxies without SNe; and unfortunately, both these types of systems are also expected to be rare in our dataset.
Therefore, to prepare a training set with boosted representation of systems that we expect to be more challenging to classify, we also use \texttt{deeplenstronomy} to inject lensed source galaxies and non-lensed supernovae into a fraction of the negative-class images.

\begin{figure*}
\vspace{3cm}
\hspace{-1.9cm}
\includegraphics[width=\textwidth,bb=0 0 1200 1000]{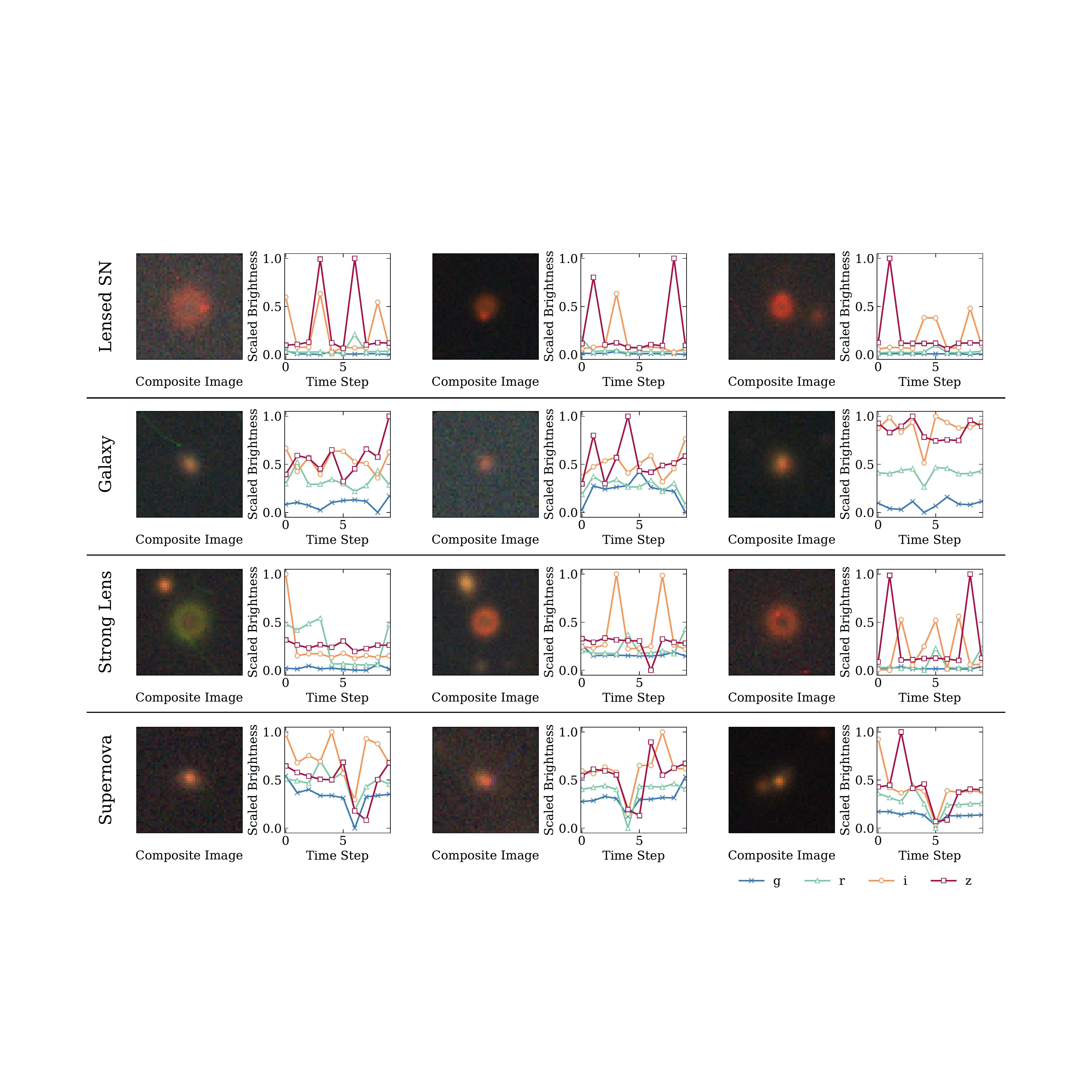}
\vspace{-4cm}
\caption{Examples of systems from our training dataset. 
The composite image is an RGB visualization of the averaged $gri$ images and the scaled brightnesses are the values extracted from the $g$ (blue ``$\times$''), $r$ (green triangles), $i$ (orange circles), $z$ (red squares) images at each time step in the time series image set using the aperture method presented in DZ1.}
\label{fig:dataset}
\end{figure*}

The process of injecting simulated light sources into real time series image sets has multiple benefits.
The training dataset includes all types of astronomical systems that the ZipperNet will classify because it is chosen from the total dataset.
Also, the properties of the simulated source galaxies and SNe are drawn from real data, maintaining all inherent physical correlations.
We join the DES Year 3 Gold catalog and DES Year 1 morphological catalog \citep[]{morphology} to obtain a sample of $\sim100,000$ galaxies from which we draw parameter values for simulations.
The simulated source galaxies are modeled with Sersic light profiles that have a color-independent ellipticity, a Sersic profile index, a band-wise half-light radius, a band-wise magnitude, and a photometric redshift --- all measured within DES pipelines.
As in DZ1, the injected SNe were simulated using public rest frame SN spectral energy distributions \citep[]{sntemplates} available in \texttt{deeplenstronomy}, which redshifts the distribution and calculates the observed magnitude in each band.
The injected SNe reach peak brightness within the interval of 20 days before the first observation and 20 days after the final observation: the dataset contains falling-only ($\sim15\%$), rising-only ($\sim15\%$), and complete lightcurves ($\sim70\%$).

To calculate the lensing effects of the real galaxy on the simulated source light, we use the measured photometric redshift of the lens galaxy, select an Einstein radius at random from the interval $[0.4\arcsec, 1.8\arcsec]$, and model the mass distribution of the lens as a singular isothermal ellipsoid following similar approaches in the literature \citep{thetae1}.
For simplicity, the mass distribution shares the measured center position and ellipticity values with the light from the real lens galaxy.
This simplification is not expected to greatly affect performance because these parameters are expected to be positively correlated.
From the mass profile, we calculate the lensed positions of the source galaxy and LSN, as well as account for the time delays of the separate SN images.
The output of the \texttt{deeplenstronomy} simulation are time series image sets with three kinds of objects added to real DES images -- LSNe, lensed source galaxies, and non-lensed SNe.

In total, 25\% of the 76,203 time series image sets placed aside for training are injected with an LSN Ia and 25\% are injected with a lensed core-collapse SN (LSN CC) to construct the positive class.
Also, 16.5\% of the training time series image sets are left untouched, 16.5\% are injected with a galaxy-galaxy strong lens, 8.25\% are injected with a SN Ia, and 8.25\% are injected with a SN CC.
The positive and negative training classes are equal in total number to maintain a balanced dataset throughout training.
We describe the details of the training in Section \ref{sec:training}, but it is worth noting here that given our choice of loss function, balancing the classes is essential to prevent class representation biasing the learned feature representation.
The remainder of this subsection describes this simulation-injection process in detail.
Examples of objects in the training dataset are collected in Figure \ref{fig:dataset}.

\subsection{Preprocessing}
\label{sec:preprocessing} 
Before we train the ZipperNet and apply it to the observed dataset, we apply a series of standardization steps.
We first truncate the time series image sets to 10 ``time steps'' in each band.
A time step refers to a single exposure in the sequence of observations; in the DES SN fields, a time step is approximately 6-7 days.
If an image set contains more time steps, we separate it into multiple 10-time step sequences: time steps 1-10 are a single sequence; time steps 2-11 are a second sequence; etc. 
Then, for each 10-step image sequence, we extract the total brightness as a function of time using the background-subtracted aperture technique presented in DZ1 with an aperture radius of 15 pixels.
Importantly, when extracting the total brightness, the zeropoint of the image is not used to maintain independence from all non-image data products.
This choice produces noise-dominated extracted brightness lightcurves, such as those in Figure \ref{fig:dataset}, though it is shown in the remainder of the analysis that the ZipperNet can still identify the temporal signatures of LSNe within the noise.

Next, we average the images within each band to obtain a single image in each band for the 10-step image sequence.
Finally, we scale the pixel values of the averaged images and the extracted brightness values linearly to range 0 to 1 on a per-example basis.
The resulting input to the ZipperNet is two different kinds of data -- 1) a scaled image in each of the $griz$ bands as a $4\times45\times45$-element array and 2) a scaled 10-step lightcurve in each of the $griz$ bands as a $4\times10$-element array.
After processing the training dataset into 10-step sequences and downsampling to maintain equal representation of the positive and negative classes, we have a total of 1,000,012 training examples.
We split these examples into 90\% training and 10\% validation datasets.

\subsection{ZipperNet}

\begin{table}
    \centering
    \begin{tabular}{|l|l|} \toprule
     \bf{Layer}    &  \bf{Specifications} \\ \midrule
     \textbf{conv1}$^\dagger$   &  Conv2D --- ($k$: 10, $p$: 2, $s$: 1) --- (4 $\rightarrow$ 16) \\ 
     \textbf{maxpool}   & MaxPool2D ($k$: 2) \\ 
     \textbf{conv2}$^\dagger$   &  Conv2D --- ($k$: 5, $p$: 2, $s$: 1) --- (16 $\rightarrow$ 32) \\ 
     \textbf{maxpool}   & MaxPool2D ($k$: 2) \\ 
     \textbf{conv3}$^\dagger$   &  Conv2D --- ($k$: 3, $p$: 2, $s$: 1) --- (32 $\rightarrow$ 64) \\ 
     \textbf{maxpool}   & MaxPool2D ($k$: 2) \\ 
     \textbf{flatten}   & Reshape (12$\times$12$\times$64 $\rightarrow$ 9216$\times$1) \\
     \textbf{fc1}$^\dagger$   & Fully-Connected (9216 $\rightarrow$ 408) \\ 
     \textbf{fc2}$^\dagger$   & Fully-Connected (408 $\rightarrow$ 25) \\ \midrule
     
     \textbf{lstm1}  & LSTM ($h$: 128) \\
     \textbf{lstm2}  &  LSTM ($h$: 128) \\
     \textbf{lstm3}  &  LSTM ($h$: 128) \\
     \textbf{fc3}$^\dagger$   & Fully-Connected (128 $\rightarrow$ 50) \\ \midrule
     
     \textbf{concat} & Concatenate \textbf{fc2} and \textbf{fc3} Outputs \\ \midrule
     
     \textbf{fc4}$^\dagger$   & Fully-Connected (75 $\rightarrow$ 6) \\ 
     \textbf{fc5}$^\ddagger$   & Fully-Connected (6 $\rightarrow$ 2) \\  \bottomrule
    \end{tabular}
    \caption{ZipperNet layer specifications. 
    We adopt the following shorthand: kernel size ($k$), padding ($p$), stride ($s$), and hidden units ($h$). 
    Arrows indicate the change in the size of the data representation as it is passed through the layer.  
    ``$^\dagger$'' indicates a Rectified Linear Unit (ReLU) activation function. 
    ``$^\ddagger$'' indicates a LogSoftmax activation function. 
    In total, our model contains 4,148,225 trainable parameters.
    \label{tab:network}}
\end{table}

The two-branch architecture of ZipperNet was first presented and validated in DZ1, and we summarize here.
One branch receives scaled, time-averaged images in each band as inputs to a block that extracts convolutional features.
The other branch receives scaled extracted brightness-time series as inputs to a block that extracts sequence features.
The outputs from the feature-extraction blocks are flattened and concatenated.
A series of fully connected layers then weights and condenses the concatenated feature representation to produce an output score that the input system contains a LSN.
The ZipperNet used in this paper is similar to Figure 2 of DZ1, and the exact hyperparameter settings for this analysis are presented in Table \ref{tab:network}.

We performed a full hyperparameter optimization of the architecture and learning algorithm using the validation dataset.
Small changes to hyperparameter settings from the prototype ZipperNet in DZ1 reflect a specialization for the real DES images used in the training data.
We find that the addition of another convolutional layer, the addition of another long short-term memory (LSTM) layer, minor tweaks to convolutional layer kernel and stride settings, and the removal of dropout layers leads to boosted performance.
The selected settings for the learning algorithm are presented in Section \ref{sec:training}.

\begin{figure*}
    \centering
    \includegraphics[width=0.49\textwidth, bb=0 0 500 350]{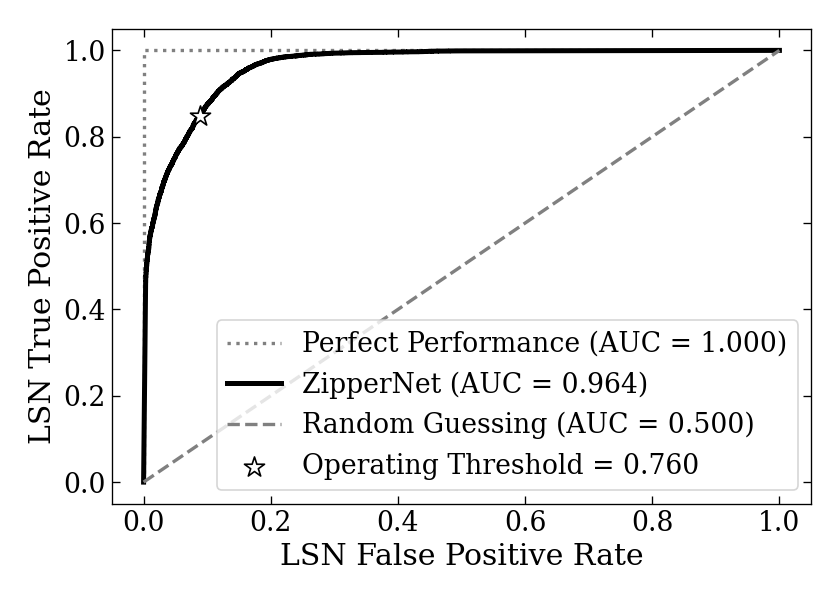}
    \includegraphics[width=0.49\textwidth, bb=0 0 460 330]{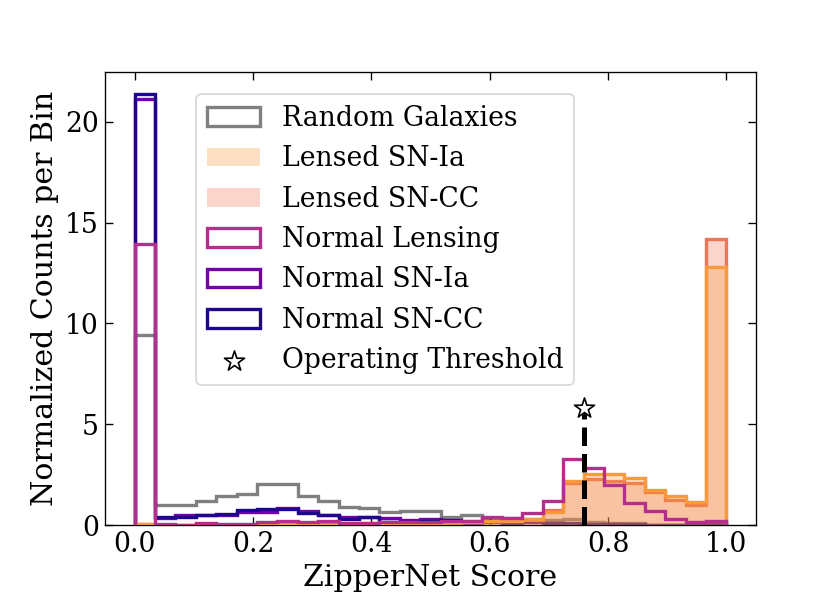}
    \caption{\textit{Left:} 
    A Receiver Operating Characteristic (ROC) curve showing the LSN true positive rate and LSN false positive rate for all possible values of the ZipperNet operating threshold. 
    The operating threshold of 0.760 is chosen to minimize the false positive rate to the point immediately prior to the true positive rate declining rapidly. 
    \textit{Right:} Histograms of the scaled ZipperNet scores for each class in the validation dataset. 
    The selected operating threshold limits false positives from all systems in the negative class while keeping the majority of the positive class.}
    \label{fig:performance}
\end{figure*}

\subsection{Training}
\label{sec:training}

To train the ZipperNet, we implemented a distributed setup on five computers (2 machines with Intel 3.2 GHz processors and 256 GB RAM, 1 machine with an AMD 2.2 GHz processor and 512 GB RAM, and 2 machines with IntelX 2.6 GHz processors and 768 GB RAM) on the DES cluster at Fermilab.
The training dataset was split into five equal chunks -- each placed on an independent computer.
On each computer, we instantiated a ZipperNet and initialized the weights at the same randomly selected values.
We then begin passing the chunks of training data through the ZipperNet instances on each of the five computers.
At regular intervals (every 1/15 of a chunk), we collect the parameters of each of the five ZipperNet instances and average the values of the parameters.
Mathematically, the averaging operation is equivalent to the weights being updated by normal training, provided the learning rate is scaled by the number of network instances.
Within this setup, we use a batch size of five examples and use stochastic gradient descent with a Nesterov momentum coefficient of 0.9, a constant learning rate of 0.001, and categorical cross-entropy loss to update the weights at each training step.
We refer to the exhaustion of all data in a chunk as a ``training iteration'' and cycle back to the beginning of the chunk once the data has all been passed through the network instance.
We allow training to continue for five training iterations and reach a final validation set accuracy of 93.0\%.
This raw accuracy is dependent on the representations of the different types of negative examples in the validation dataset.
In Section \ref{sec:results}, we assess the performance using physically meaningful metrics.

\subsection{Candidate Selection Criteria}

The output of the trained ZipperNet on an input (pair of an averaged image and a lightcurve) is a score with a value typically between -100.0 and 50.0.
Based on the minimum and maximum values of this range in our validation dataset, we linearly scale the ZipperNet output scores to the range $[0.0, 1.0]$, such that they are similar to probabilities.
Next, we select a threshold ZipperNet score above which we include the candidate system in our final sample and below which we exclude the candidate system.
We select this threshold by iterating through possible threshold values and analyzing the fraction of LSNe that scored higher than the threshold compared to the fraction of galaxies that scored higher than the threshold.
The left panel of Figure \ref{fig:performance} shows the attainable values of these quantities for different thresholds.
We expect galaxies to be the largest background: the number of galaxies in a given area of sky is orders of magnitude higher than the number of SL or SNe.
Therefore, we select the threshold by reducing the fraction of galaxies scored higher than the threshold to the lowest value before the fraction of LSNe scored higher than the threshold starts to decline rapidly.
Based on this analysis, we select an operating threshold for the scaled ZipperNet scores of 0.76.
This threshold value is contextualized with the ZipperNet scores for the systems in our validation dataset in the right panel of Figure \ref{fig:performance}.

\begin{figure}
    \centering
    \includegraphics[width=\columnwidth, bb=0 0 480 330]{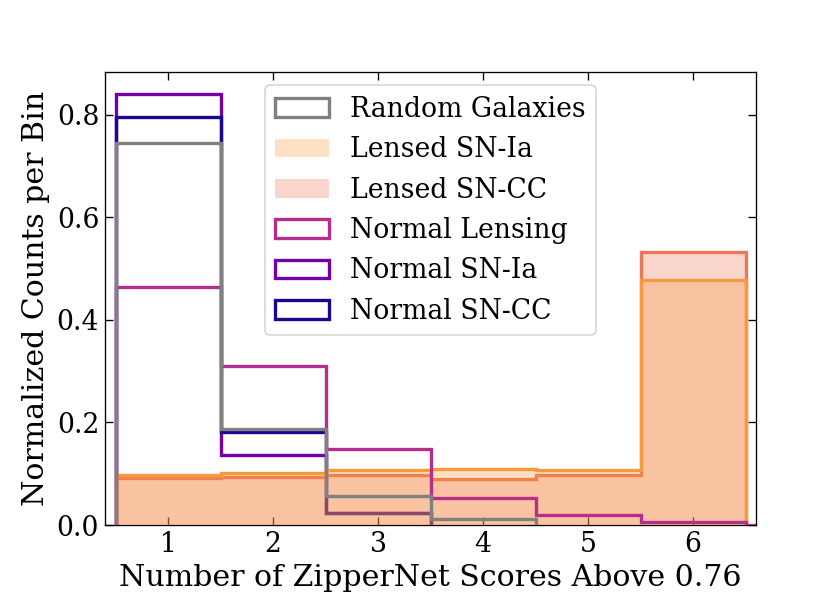}
    \caption{The number of time series image set subsequences scored above the ZipperNet threshold for each type of object in our validation dataset. 
    On average, LSNe time series image sets are scored above the ZipperNet threshold in a higher fraction of their subsequences than all types of negative examples.}
    \label{fig:detections}
\end{figure}

We develop a final selection criterion to narrow the sample of candidate systems selected by ZipperNet.
We leverage the aspect of our data processing from Section \ref{sec:preprocessing} in which time series image sets with more than 10 epochs are split into 10-epoch subsequences, which are then classified independently by ZipperNet.
In analyzing the ZipperNet classifications made on all subsequences of a time series image set, we find that LSNe are more likely than galaxies to have multiple detections.
This relationship is illustrated in Figure \ref{fig:detections} using our validation dataset, which we use as motivation to develop a criterion on the aggregate detections in a time series image set.
Importantly, the total length of the time series image sets in our training and validation data was not required to match the real data as a result of our preprocessing methods, so it would be inaccurate to set a strict requirement on the number of ZipperNet detections (score above the threshold) based on the validation dataset.
Rather, to put the validation dataset and the real data on the same footing, we set a requirement on the ratio of number of detections to number of subsequences.
Therefore, we select the threshold for this ratio such that the false positive rates are minimized to the point where human inspection of the final sample becomes feasible.
We choose to require at least 60\% of the subsequences to have a ZipperNet score above 0.76 for the candidate system to be included in our final sample of candidate LSNe.
The 60\% threshold and the 0.76 ZipperNet score threshold were determined simultaneously by computing the LSN recall and galaxy false positive rate at all possible values.

\section{Results}
\label{sec:results}

\subsection{Performance Metrics}
\label{sec:metrics}

To evaluate the performance of the fully trained ZipperNet, we define quantities and metrics of interest and compute them on the validation dataset.
We introduce two terms that describe classification score thresholds: ``classified as an LSN'' means the candidate system had a ZipperNet score greater than the threshold in at least 60\% of subsequences; and ``classified as background'' means the candidate system had a ZipperNet score greater than the threshold in fewer than 60\% of subsequences.
We define the following terms regarding metrics based on the threshold score: 
\begin{itemize}
    \item a \textit{true positive} (TP) is an LSN, and it is classified as an LSN;
    \item a \textit{false positive} (FP) is a galaxy, galaxy-galaxy lens, or unlensed SN, and it is classified as a LSN; 
    \item a \textit{true negative} (TN) is a galaxy, galaxy-galaxy lens, or unlensed SN, and it is classified as background; and
    \item a \textit{false negative} (FN) is a LSN, and it is classified as background.
\end{itemize}
Using these quantities, common metrics like accuracy are straightforward to compute; however, those metrics are misleading due to the boosted representation of rare physical systems in our training and validation datasets.
We instead focus on class-specific metrics that carry physical meaning and are robust against the class representation in the validation dataset: the LSN recall is
\begin{align}
    \textrm{LSN Recall} = \textrm{TP / (TP + FN)};
    \label{eq:lsn_recall}
\end{align}
the LSN-type-specific recall is
\begin{align}
        \textrm{LSN}_\textrm{type} \textrm{ Recall} = \textrm{TP}_{\textrm{type}} \textrm{ / }  (\textrm{TP}_{\textrm{type}} + \textrm{FN}_{\textrm{type}}),
        \label{eq:lsn_specific_recall}
\end{align}
where type  is ``Ia''  or ``CC''; and the false positive rate for each type of negative class is
\begin{align}
        \textrm{FPR}_\textrm{type} = \textrm{FP}_{\textrm{type}} \textrm{ / } (\textrm{TN}_{\textrm{type}} + \textrm{FN}_{\textrm{type}}),
        \label{eq:fpr}
\end{align}
where type is ``Galaxy,'' ``SL,'' ``SN-Ia,'' ``SN-CC.'' 
The values of these metrics are collected in Table \ref{tab:metrics} for ZipperNet alone and for the combination of ZipperNet with our final sample-selection criterion. 

\begin{table}
    \centering
    \begin{tabular}{|l|c|c|c|} \toprule
    \bf{Metric} & \bf{ZipperNet} & \bf{+ Final} & \bf{Equation} \\ 
    \midrule
    \bf{LSN Recall} & \bf{0.8447} & \bf{0.6113} & Eq. \ref{eq:lsn_recall} \\
    LSN$_\textrm{Ia}$ Recall & 0.8426 & 0.5949 & Eq. \ref{eq:lsn_specific_recall} \\
    LSN$_\textrm{CC}$ Recall & 0.8468 & 0.6273 & Eq. \ref{eq:lsn_specific_recall} \\
    \bf{FPR$_\textrm{Galaxy}$} & \bf{0.0157} & \bf{0.0002} & Eq. \ref{eq:fpr} \\
    FPR$_\textrm{SL}$ & 0.2448 & 0.0046 & Eq. \ref{eq:fpr} \\
    FPR$_\textrm{SN-Ia}$ & 0.0049 & 0.0001$^\dagger$ & Eq. \ref{eq:fpr} \\
    FPR$_\textrm{SN-CC}$ & 0.0046 & 0.0001$^\dagger$ & Eq. \ref{eq:fpr} \\
    \bottomrule
    \end{tabular}
    \caption{
    Metrics for evaluating the performance of ZipperNet and our final sample selection that are robust against the class representations of the validation dataset. 
    All metrics are defined in Section \ref{sec:metrics}. 
    ``$\dagger$'' indicates the use of an upper limit on the metric value resulting from limited statistical precision.}
    \label{tab:metrics}
\end{table}

There are a few key results from these metrics worth highlighting.
The ZipperNet LSN recall indicates that approximately 84\% of all LSN in the validation dataset are scored above the operating threshold.
The ZipperNet galaxy false positive rate indicates that roughly 1.5\% of galaxies will be scored above our operating threshold and erroneously populate our candidate sample.
By itself, the ZipperNet is a powerful classifier, but the minimized galaxy false positive rate is still large enough where the resulting candidate sample would be too large for visual inspection.
With the addition of the selection criterion on the number of ZipperNet detections for each constituent subsequence, the performance is boosted.
Critically, the final galaxy false positive rate is reduced, facilitating visual inspection of the full final candidate sample.
This stricter selection has the consequence of reducing the final LSN recall.
However, most of the removed LSNe are those that peak before or after the window of observations.

\begin{figure*}
    \centering
    \includegraphics[trim=4.2cm 17.5cm 3.5cm 17.5cm,clip,width=0.82\textwidth]{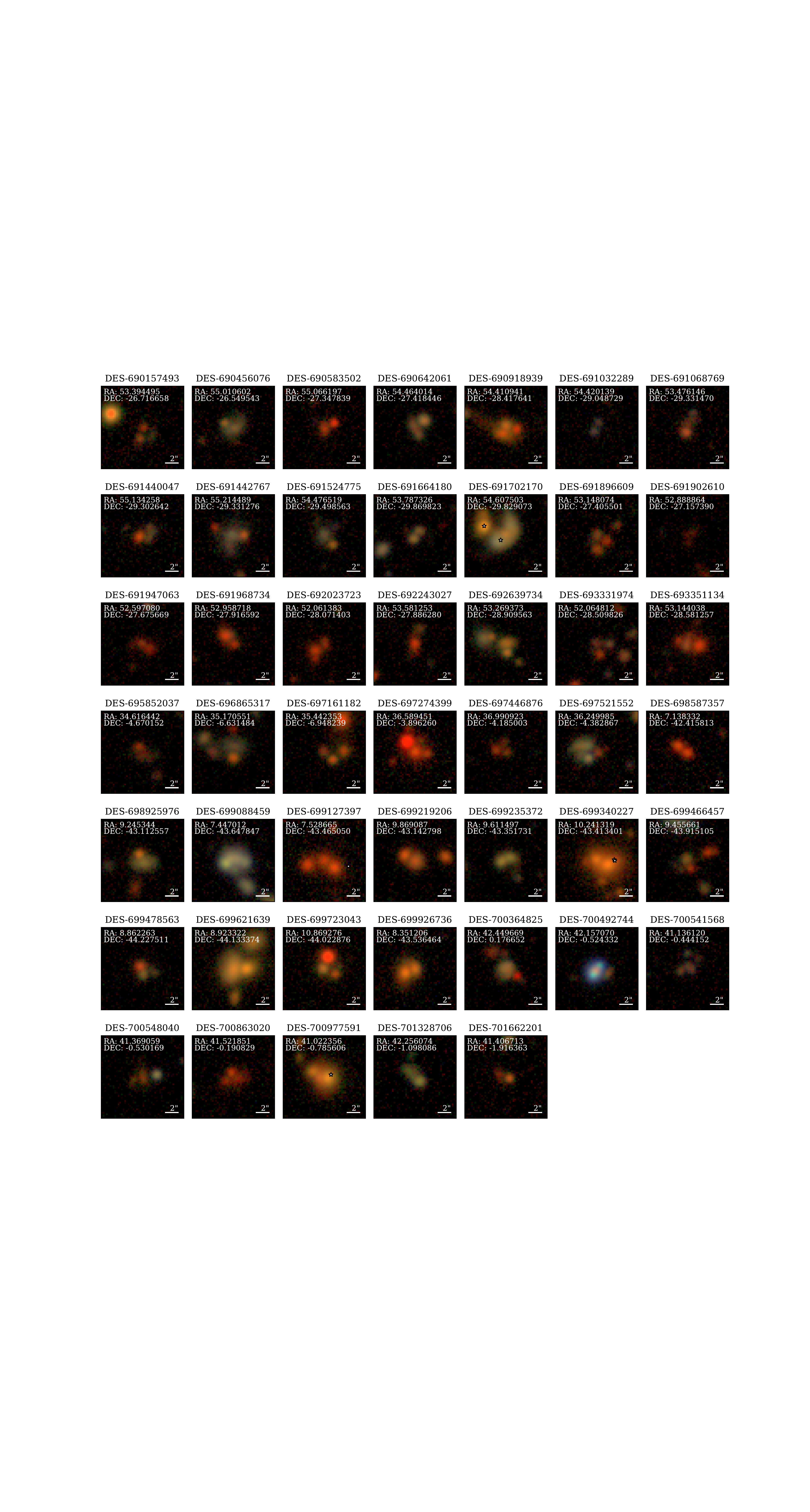}
    \caption{
    Candidate systems detected by ZipperNet that showed evidence of lensing but do not show SNe-like variability in their lightcurves. 
    The properties of these candidates are collected in Table \ref{tab:candidates}.
    Difference imaging detections from the DES SN group are shown with white star markers.}
    \label{fig:almosts}
\end{figure*}

\subsection{Searching the DES SN Fields}

Applying our trained ZipperNet and additional selection criterion to the DES SN field data produces 2245 candidate LSNe, approximately half of which had ZipperNet detections in multiple years of DES data.
We expect the majority of these systems to have resolvable features based on two aspects of the analysis.
First, these 2245 candidate LSNe were identified in the magnitude-limited sample of the DES galaxies, leading to a tendency for low redshift, nearby galaxies to be more highly represented than high redshift, distant galaxies.
Second, based on the physical selection function of the ZipperNet on this dataset (shown in Appendix \ref{app:selection}), LSNe in systems with large Einstein radii and better seeing are more likely to be recalled. Therefore, because the majority of the systems in this candidate sample should have resolvable features, human visual inspection becomes a viable approach for identifying the most interesting candidate LSNe.

A team of strong lensing experts within DES inspected 6-year coadded, color-composite images of the 2245 candidate LSNe systems to search for lensing, similar to how precursor strong lensing searches have been carried out.
The team assigned all objects a score using the following system:
\begin{itemize}
  \item[\bf{1:}] The detection is an image artifact, such as a diffraction spike or contamination from a bright foreground star;
  \item[\bf{2:}] There is a single object, such as a galaxy or star;
  \item[\bf{3:}] There are multiple objects with no evidence of lensing, such as SNe or clusters of galaxies;
  \item[\bf{4:}] There are multiple objects with evidence of lensing.
\end{itemize}
Using this system and the median score for each object, the team of inspectors identified 522, 802, 871, and 50 objects with scores ``1,'' ``2,'' ``3,'' ``4,'' respectively.
For the 50 systems with evidence of lensing, we extracted aperture lightcurves for each object in each system from the DES single-epoch images.
Three systems from the 50 systems with evidence of lensing were identified to have SN-like time variability, and we upgraded their overall score to a ``5.''
Candidate systems scored as a ``4'' or ``5'' are presented in Figures \ref{fig:almosts} and \ref{fig:candidate_lsne}, respectively, and have their properties collected in Table \ref{tab:candidates}.

The 50 candidate systems scored at or above a ``4'' found in this analysis show evidence of lensing in their images.
There are still non-lenses in this sample: for example, DES-700492744 is a high-proper motion white dwarf appearing as a red object between two blue point sources; nevertheless, we include all systems labeled as interesting by the labeling team for completeness.
Some of these systems also show evidence of point sources within the lensing configurations: there are nearly circular objects positioned within the lensing configuration.
Going further, we analyze the time variability of the candidate systems by extracting 5-year background-subtracted lightcurves for each source in the images.
The objects scored as a ``5'' show evidence of SN-like time variability: a short rise followed by a steady decay in brightness over the course of approximately one month as shown by Figure \ref{fig:candidate_lsne}.
The objects scored as a ``4'' do not show this temporal behavior, however the possibility remains that some of the objects scored as a ``4'' are strongly lensed systems and potentially house a lensed quasar.
Section \ref{sec:final_cands} contains a detailed presentation of the three objects scored as a ``5''.

Lastly, we cross-match the 2,245 ZipperNet-identified systems  with the systems identified during the DES 5-year photometric SN Ia cosmology analysis \citep{dessn_photo}.
In \citet{dessn_photo}, difference imaging \citep{diffimg} identified 31,636 transients and SALT-II SN Ia lightcurve fitting \citep{salt2} identified 2,381 single-season SNe from that sample of transients.
The SNe selected by lightcurve fitting are more likely to be SNe Ia than SNe CC, and most SNe CC in the total sample are also excluded by the fitting.
Furthermore, this selection procedure searches for normal SNe Ia and is not adapted for possible changes in the light-curves from the lensing.
In total, there is an overlap (using a 5\arcsec radius) of 104 systems amongst the ZipperNet sample and the DES SN analysis transient sample.
All but four overlapping systems --- DES-691702170, DES-699127397, DES-699340227, and DES-700977591 --- were scored as either a ``2'' or a ``3'' by the labeling team, indicating no convincing evidence for lensing.
The locations of the detected transients are marked in Figure \ref{fig:almosts}.
Only the transient in DES-699127397 passed the SALT-II SN Ia lightcurve fitting.
The difference-imaging detections in DES-699340227 and DES-700977591 appear to be spurious detections due to image subtractions errors.
Lastly, while the transients detected in DES-691702170 and DES-699127397 are likely SNe, these systems do not appear to be lenses and likely should have received lower grades from the labeling team;
DES-691702170 lacks an obvious lensing galaxy and the positions of the galaxies in DES-699127397 are more likely a cluster of galaxies than multiple images of the same background galaxies due to their asymmetric alignment.

Based on the SN FPRs in Table \ref{tab:metrics}, this overlap is consistent with the expected ZipperNet SN background.
The three systems scored as a 5 by the visual inspection team, indicating the presence of both lensing and SN-like temporal behavior, were not included in the overlapping sample.
We believe the faintness of the SNe or foreground contamination for the lensing galaxy may have contributed to the non-detection from difference imaging, though a full understanding of this discrepancy is beyond the focus of our analysis.

\subsection{Final LSN Candidates}
\label{sec:final_cands}

\begin{figure*}
    \centering
    \includegraphics[width=\textwidth]{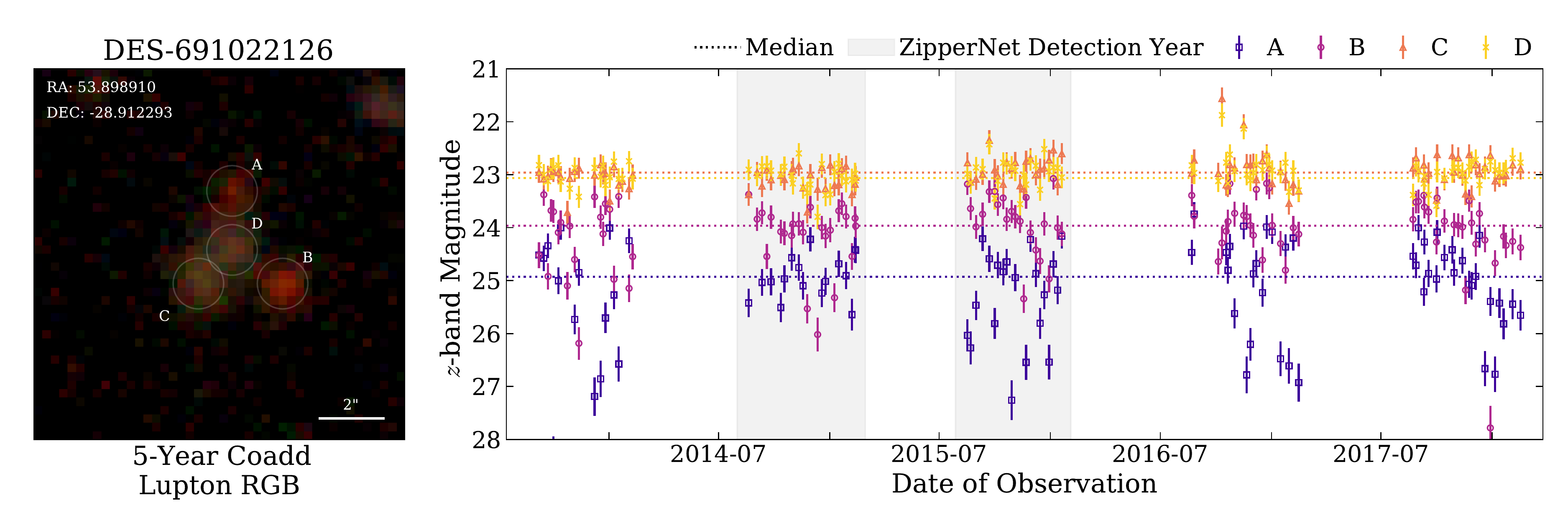}
    \includegraphics[width=\textwidth]{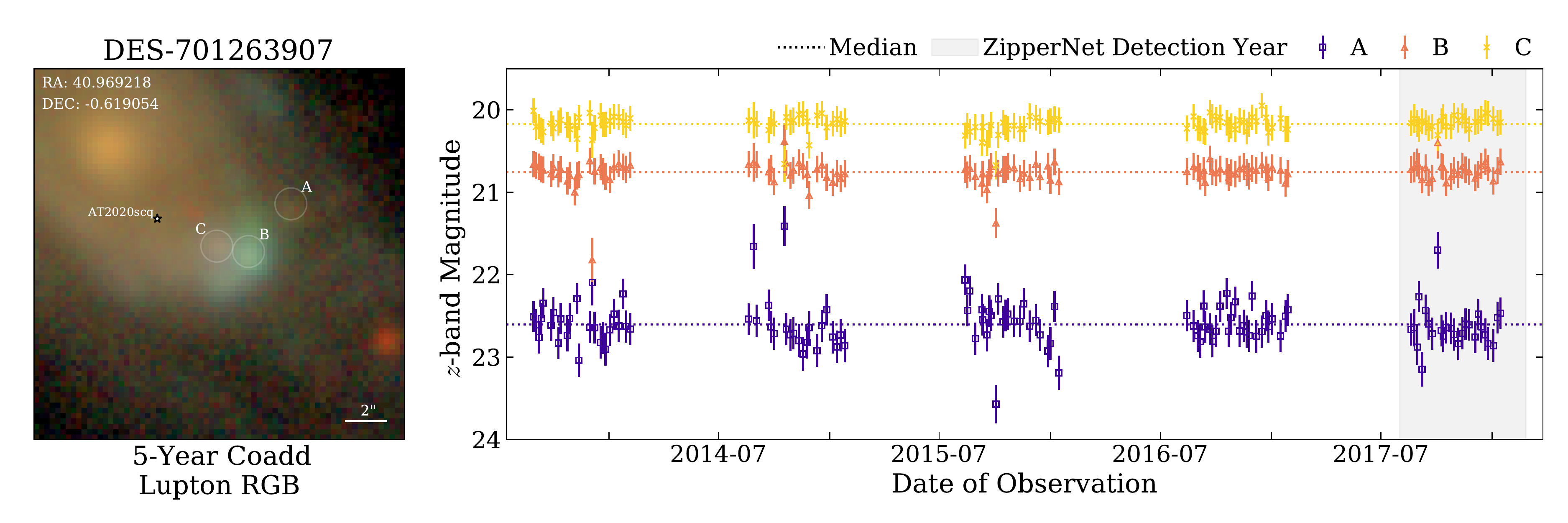}
    \includegraphics[width=\textwidth]{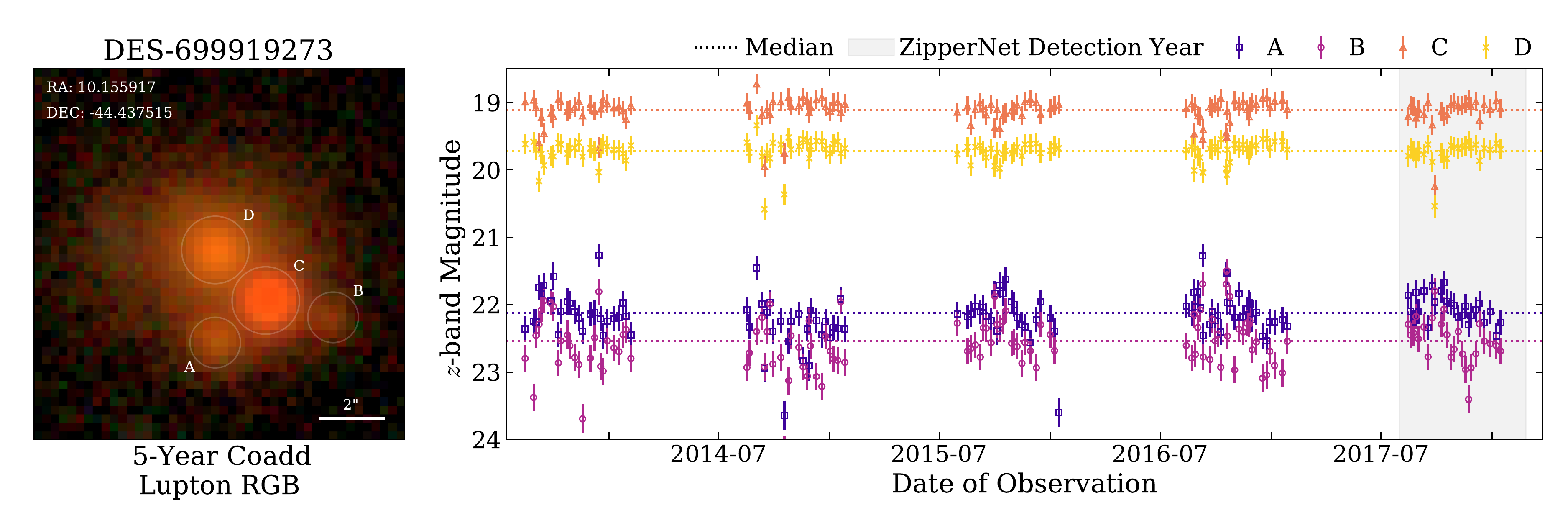}
    \caption{The candidate LSNe identified by ZipperNet and human visual inspection. 
    The aperture used to extract the magnitude measurement from each source is show and annotated on the coadded image.
    The properties of the candidates are collected in Table \ref{tab:candidates}.}
    \label{fig:candidate_lsne}
\end{figure*}

The three most interesting systems identified by the ZipperNet and subsequent human visual inspection are DES-691022126, DES-701263907, and DES-699919273.
We present 5-year color-composite coadded images of these systems and extract lightcurves for each object of interest within them in Figure \ref{fig:candidate_lsne}.
From the lightcurves, the five observing seasons of the DES SN program are easily distinguishable, and we refer to each observing season as ``Y1'' through ``Y5.''
We extract the lightcurves from the single-epoch images by summing the pixels in the aperture displayed in the coadded image, subtracting the sky background measured by DES, and converting to a magnitude using the zeropoint measured by DES.
Importantly, the magnitudes are the combination of all objects within the aperture, so for example an SN lightcurve will contain contamination from its host galaxy.
All estimated Einstein radii have been obtained by measuring the angular separations between objects, as opposed to a full modeling of the lensing system.
We choose to present only the $z$-band lightcurves for these visualizations for simplicity, though all four bands were assessed to identify SN-like temporal behavior.
The bluer bands such as $g$ and $r$ have larger PSFs in this dataset compared to the redder $i$ and $z$ bands, leading to noisier aperture photometry measurements.
Furthermore, LSNe are likely to be at high redshifts, leading to a tendency for LSN temporal signatures to be most visible in the redder bands.

DES-691022126 is a system of four objects labeled in the top panel of Figure \ref{fig:candidate_lsne} as A, B, C, and D.
We interpret objects C and D to be galaxies based on their constant brightness over time.
Objects A and B are much redder, and display a greater degree of brightness variability when looking at the typical size of the magnitude error bars compared to the 5-year median $z$-band magnitude for each object.
Furthermore, the lightcurves for objects A and B both contain a period of linear decline in magnitude on month timescales: object A in Y5 and object B in Y3.
ZipperNet detected the system in Y2 and Y3, but not in Y5.
We believe it detected the linear decline of object B in Y3 and that perhaps object C contained light from a SN between Y2 and Y3 for which ZipperNet detected the beginning of.
The fact that the linear decline of object A's brightness in Y5 was not detected by ZipperNet is likely due to object A being the faintest source in the system and the selection function of ZipperNet (see the bottom right panel of Figure \ref{app:selection}).
The SN-like lightcurve features that are shared between objects A and B, when combined with the evidence for lensing with an Einstein radius of approximately 1.7\arcsec, support the claim of the system as a LSN.

By comparison, DES-701263907 is a much more complicated system shown in the middle panel of Figure \ref{fig:candidate_lsne}. 
A large foreground galaxy (SIMBAD source LEDA 135660) at redshift $0.03$ dominates the image. 
Object B (SIMBAD source SDSS J024352.54-003708.4) is cataloged as a galaxy also at redshift  $0.03$, but may also be a dense, star-forming region based on its blue color.
This dense area, combined with the gravitational potential of LEDA 135660 itself would have a large lensing cross section, increasing the likelihood that background objects would be lensed.
Because the lightcurve extraction method used in the lightcurves of Figure \ref{fig:candidate_lsne} does not subtract the effect of the host galaxy, the variability of these objects cannot be assessed without difference imaging techniques outside the scope of this paper.
Nonetheless, we identify object A as the most variable source in the system given the foreground contamination.
In the image cutout for DES-701263907, we also note the location of an SN detected in September of 2020 (AT2020scq).
It is possible that object B acts as a primary lensing galaxy, object A is an LSN identified by ZipperNet in 2018, and AT2020scq is a second appearance of object A delayed by approximately two years.
Given the large foreground galaxies at redshift $0.03$ and a potential Einstein radius of $\approx3.0$\arcsec, this time delay would be consistent with an LSN.

Lastly, DES-699919273 is another four-object system that we enumerate as A, B, C, and D in the bottom panel of Figure \ref{fig:candidate_lsne}.
We interpret object C as the lensing galaxy, object D as an image of the source galaxy without an SN, and objects A and B as images of the source galaxy, where an SN was present at some point during DES observations.
The Einstein radius for this system is $\approx2.1$\arcsec.
Particularly, we note a linear decline in $z$-band magnitude for object A in Y3 and a nearly identical linear decline in $z$-band magnitude for object B in Y5.
ZipperNet detected the linear decline in Y5, but had no such detection in Y3.
We interpret this event as another manifestation of the less-than-perfect recall of the classifier.
Nonetheless, the SN-like temporal signal appearing in two of the images within a lensing geometry is evidence for the presence of an LSN.

\section{Discussion}
\label{sec:discussion}

The method presented in this analysis contains a few areas where improvements could increase the LSN recall while decreasing the false positive rate.
One such change is to add centroiding to account for sub-pixel-level shifts in position prior to stacking and averaging the images.
While the offsets are small, misalignment at the $\approx0.25$\arcsec\ scale can cause image-based features to become less sharp and harder for a convolutional layer to identify.
When stacking the images, it may also boost performance to only include the images with high image quality (e.g. seeing above some quality threshold and~/~or cloudiness below some quality threshold): this would ensure that the ability to resolve features in the resulting composite image is only limited by the instrumentation.
These possibilities focus on improving the appearance of spatial features in the data to boost the ZipperNet's ability to learn relationships and are motivated by the analysis of the physical selection function of our approach, which is described in Appendix \ref{app:selection}.

The lightcurve extraction step of the data preprocessing also could be improved by discarding common artifacts such as diffraction spikes and saturated pixels to avoid contaminating the extracted brightness.
Similarly, an analysis of the clarity of features in the lightcurves as a function of the aperture radius used in the lightcurve extraction may find that a different aperture radius leads to higher performance.
It's possible that scaling the time series image sets to have a standardized mean and a standardized variance of pixel value prior to preprocessing would lead to smoother lightcurves.
Finally, the preprocessing steps used in this analysis down-selected images to standardize the cadence, though other approaches have demonstrated success with arbitrary numbers of images in the time series \citep{spacio}.
Removing the need for a standardized cadence would greatly improve the applicability of this approach to real-time LSNe identification and remove the need for images to be discarded.

It's possible that the machine learning aspect of the analysis could be improved with subtle changes to the training set.
For example, when simulating lensed systems, we made the simplifying approximation that the mass profile ellipticity was equivalent in angle and strength to the light profile.
While there is likely to be a strong correlation between the mass and light profiles, the exclusion of training set examples with different relationships between mass profile ellipticity and light profile ellipticity may bias LSN selection to systems in which these quantities are highly correlated.
We also employed a uniform distribution of Einstein radii, and it is possible that an approach such as \citet{spacio} with a physically motivated distribution could lead to improved performance.

In consideration of a real-time LSN detection pipeline, a couple changes to the methodology may improve performance.
We envision the 10-epoch time series image sets being constructed as observations are ongoing: after a new image of a system is collected, the first image in the time series is discarded and a new 10-epoch sequence is created.
There are two downsides to that approach: (1) there is an implicit requirement of 10 epochs before the trained ZipperNet can be utilized, and (2) our final selection criterion on the fraction of subsequences scored above the ZipperNet threshold requires additional epochs to create and track multiple subsequences.
With the improvements to the preprocessing discussed above, it may be possible to sufficiently boost the ZipperNet performance to the point where the additional selection criterion can be removed.
Furthermore, we did not experiment with time series image sets with fewer than 10 epochs, and it is possible that the analysis can be performed with a less strict requirement on the total number of epochs. 

With the current configuration, we have successfully reduced a catalog of 3,459,186 objects to 2,245 with our deep learning approach, and proceeded to identify 50 systems of interest through human visual inspection, three of which show some evidence of an LSN.
While we do not confirm or further characterize these three systems of interest, they all contain lensing features and the presence of point sources as found during the human visual inspection.
Full characterization would entail Scene Modeling Photometry \citep{BroutSMP} to obtain lightcurves without host galaxy contamination, photometric classification of that time-series photometry, redshift measurements for all objects in the system, and lens modeling which are beyond the scope of this search.
Because any detected LSNe would have faded by now, follow-up observations to confirm LSNe are unlikely to provide any additional information apart from redshifts.
However, several of the systems of interest were detected by ZipperNet in multiple years throughout DES operations.
Therefore, these persistent lensed systems with point sources offer interesting candidates for lensed quasar searches.
The three most interesting candidates (DES-691022126, DES-701263907, and DES-699919273) are the most likely LSNe found by our ZipperNet in the magnitude-limited 5-year DES dataset utilized in this analysis.
Given the approximate time delays and Einstein radii of the systems, spectroscopic redshifts and lens modeling could produce three independent measurements of $H_0$.

The ZipperNet architecture itself provides a new and powerful LSN identification tool going forward.
The accompanying code for this analysis \citep{zenodo_dz2} also makes the data collection, processing, simulation, training, classification, and candidate selection routines available for future analyses.
With first light from the Vera C. Rubin Observatory quickly approaching, setting up a pipeline to detect LSNe is vital for time-delay cosmography measurements.
The analysis presented here and suggested improvements provide a template for one such pipeline that would facilitate real-time detection of LSNe in short time series sequences of images without a dependence on traditional and computationally expensive image processing algorithms.

\section{Conclusion}
\label{sec:conclusion}

This analysis presents the application of a deep learning LSN detection algorithm to an observed optical survey dataset.
The algorithm utilizes a novel neural network architecture called a ZipperNet that simultaneously learns characteristic features from image and temporal data to identify LSNe in DES data.
Using a ZipperNet trained on simulated LSNe that are injected into the DES SN field data -- along with a selection criterion on the number of detections for each system -- our approach performs with an LSN recall of 61.13\% and a false positive rate of 0.02\%.
This technique identified 2,245 candidate LSN systems in the DES SN fields, and a human visual inspection found 50 systems of interest, three of which contained evidence of a time-variable lensed source.
Confirmation of these candidates of interest is left for future work, and these systems may facilitate direct measurements of $H_0$ when fully characterized.
Looking to the Rubin Observatory era, the approach developed in DZ1 and implemented on the DES SN fields here has the potential to aid in the identification of several hundred LSNe.

\section*{Acknowledgments}

R. Morgan thanks the Universities Research Association Fermilab Visiting Scholars Program for funding his work on this project. 
R. Morgan also thanks the LSSTC Data Science Fellowship Program, which is funded by LSSTC, NSF Cybertraining Grant \#1829740, the Brinson Foundation, and the Moore Foundation; his participation in the program has benefited this work. 

We acknowledge the Deep Skies Lab as a community of multi-domain experts and collaborators who’ve facilitated an environment of open discussion, idea-generation, and collaboration. This community was important for the development of this project.

This material is based upon work supported by the National Science Foundation Graduate Research Fellowship Program under Grant No. 1744555. Any opinions, findings, and conclusions or recommendations expressed in this material are those of the author(s) and do not necessarily reflect the views of the National Science Foundation.

Funding for the DES Projects has been provided by the U.S. Department of Energy, the U.S. National Science Foundation, the Ministry of Science and Education of Spain, 
the Science and Technology Facilities Council of the United Kingdom, the Higher Education Funding Council for England, the National Center for Supercomputing 
Applications at the University of Illinois at Urbana-Champaign, the Kavli Institute of Cosmological Physics at the University of Chicago, 
the Center for Cosmology and Astro-Particle Physics at the Ohio State University,
the Mitchell Institute for Fundamental Physics and Astronomy at Texas A\&M University, Financiadora de Estudos e Projetos, 
Funda{\c c}{\~a}o Carlos Chagas Filho de Amparo {\`a} Pesquisa do Estado do Rio de Janeiro, Conselho Nacional de Desenvolvimento Cient{\'i}fico e Tecnol{\'o}gico and 
the Minist{\'e}rio da Ci{\^e}ncia, Tecnologia e Inova{\c c}{\~a}o, the Deutsche Forschungsgemeinschaft and the Collaborating Institutions in the Dark Energy Survey. 

The Collaborating Institutions are Argonne National Laboratory, the University of California at Santa Cruz, the University of Cambridge, Centro de Investigaciones Energ{\'e}ticas, 
Medioambientales y Tecnol{\'o}gicas-Madrid, the University of Chicago, University College London, the DES-Brazil Consortium, the University of Edinburgh, 
the Eidgen{\"o}ssische Technische Hochschule (ETH) Z{\"u}rich, 
Fermi National Accelerator Laboratory, the University of Illinois at Urbana-Champaign, the Institut de Ci{\`e}ncies de l'Espai (IEEC/CSIC), 
the Institut de F{\'i}sica d'Altes Energies, Lawrence Berkeley National Laboratory, the Ludwig-Maximilians Universit{\"a}t M{\"u}nchen and the associated Excellence Cluster Universe, 
the University of Michigan, NSF's NOIRLab, the University of Nottingham, The Ohio State University, the University of Pennsylvania, the University of Portsmouth, 
SLAC National Accelerator Laboratory, Stanford University, the University of Sussex, Texas A\&M University, and the OzDES Membership Consortium.

Based in part on observations at Cerro Tololo Inter-American Observatory at NSF’s NOIRLab (NOIRLab Prop. ID 2012B-0001; PI: J. Frieman), which is managed by the Association of Universities for Research in Astronomy (AURA) under a cooperative agreement with the National Science Foundation.

The DES data management system is supported by the National Science Foundation under Grant Numbers AST-1138766 and AST-1536171.
The DES participants from Spanish institutions are partially supported by MICINN under grants ESP2017-89838, PGC2018-094773, PGC2018-102021, SEV-2016-0588, SEV-2016-0597, and MDM-2015-0509, some of which include ERDF funds from the European Union. IFAE is partially funded by the CERCA program of the Generalitat de Catalunya.
Research leading to these results has received funding from the European Research
Council under the European Union's Seventh Framework Program (FP7/2007-2013) including ERC grant agreements 240672, 291329, and 306478.
We  acknowledge support from the Brazilian Instituto Nacional de Ci\^encia
e Tecnologia (INCT) e-Universe (CNPq grant 465376/2014-2).

This paper has gone through internal review by the DES collaboration.
This manuscript has been authored by Fermi Research Alliance, LLC under Contract No. DE-AC02-07CH11359 with the U.S. Department of Energy, Office of Science, Office of High Energy Physics.

\software{
\texttt{astropy} \citep[]{astropy},
\texttt{deeplenstronomy} \citep[]{deeplenstronomy}, 
\texttt{lenstronomy} \citep[]{lenstronomy, lenstronomy2},
\texttt{matplotlib} \citep[]{matplotlib},
\texttt{numpy} \citep[]{numpy}, 
\texttt{pandas} \citep[]{pandas},
\texttt{PlotNeuralNet} \citep[]{network_vis}, 
\texttt{PyTorch} \citep[]{pytorch},
\texttt{Scikit-Learn} \citep[]{sklearn},
\texttt{scipy} \citep[]{scipy}.
}

\clearpage

\appendix
\numberwithin{figure}{section}
\numberwithin{table}{section}

\section{Lensed SN Physical Selection Function}

\begin{figure*}[b]
    \centering
    \includegraphics[width=0.49\textwidth]{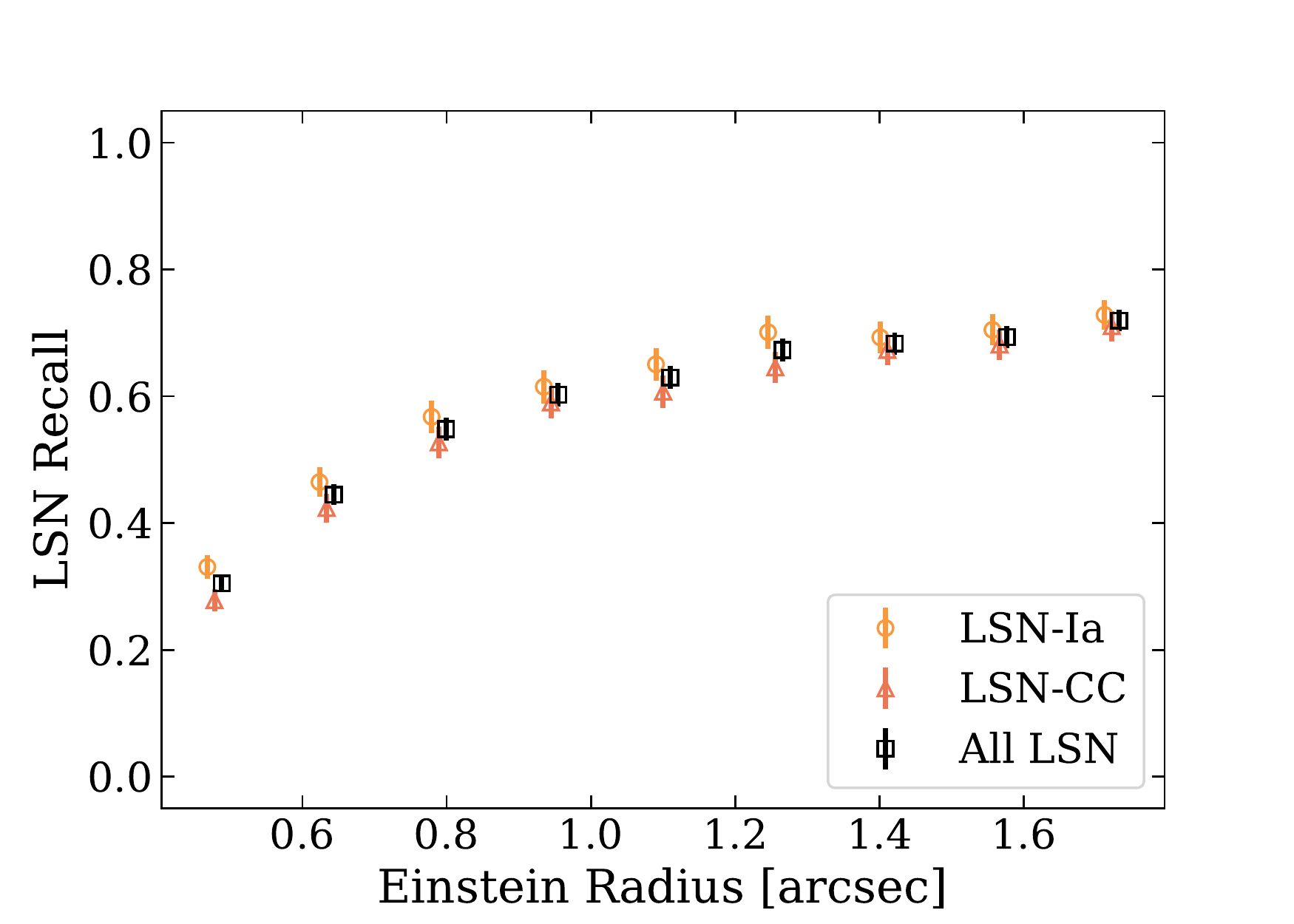}
    \includegraphics[width=0.49\textwidth]{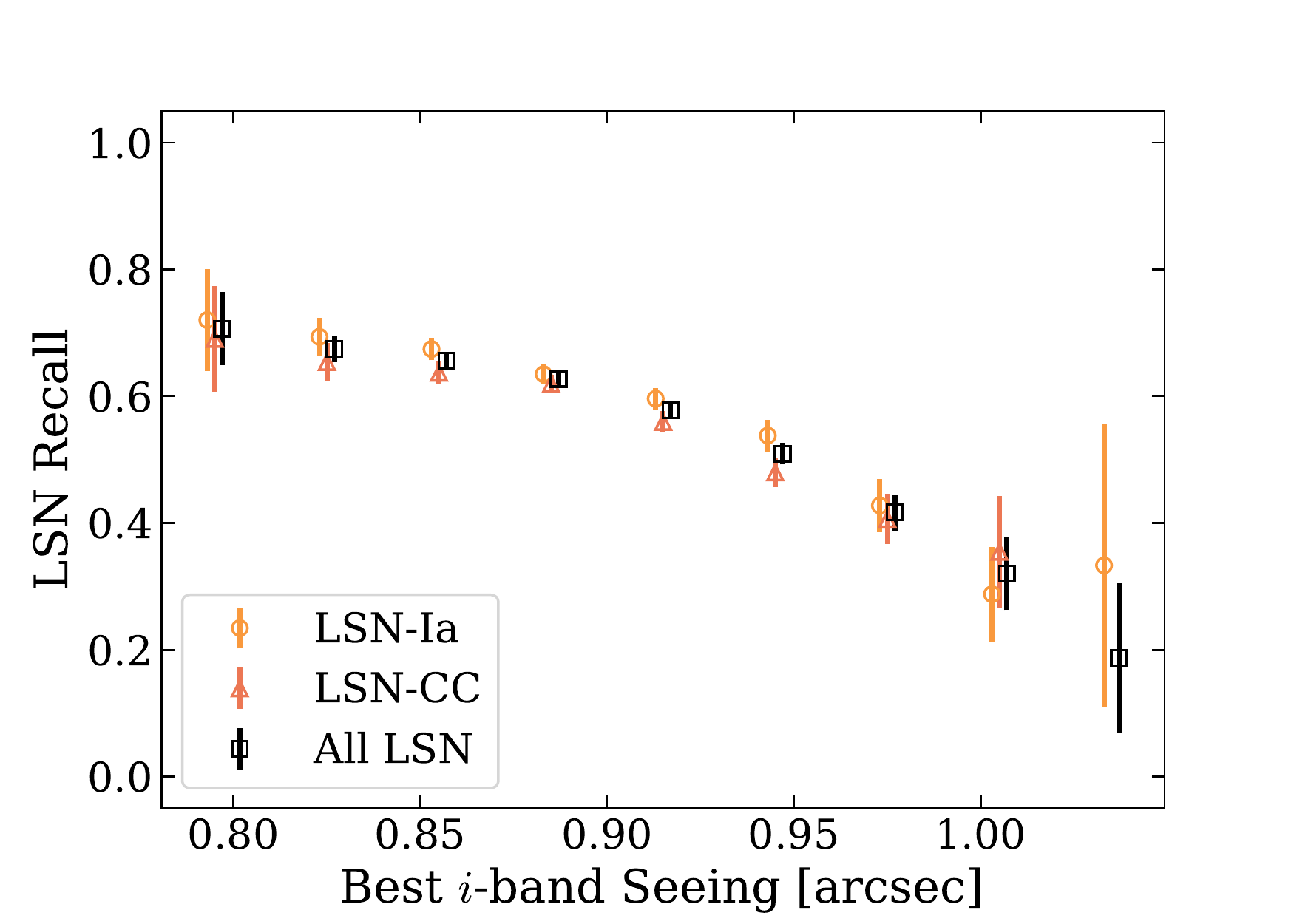}
    \includegraphics[width=0.49\textwidth]{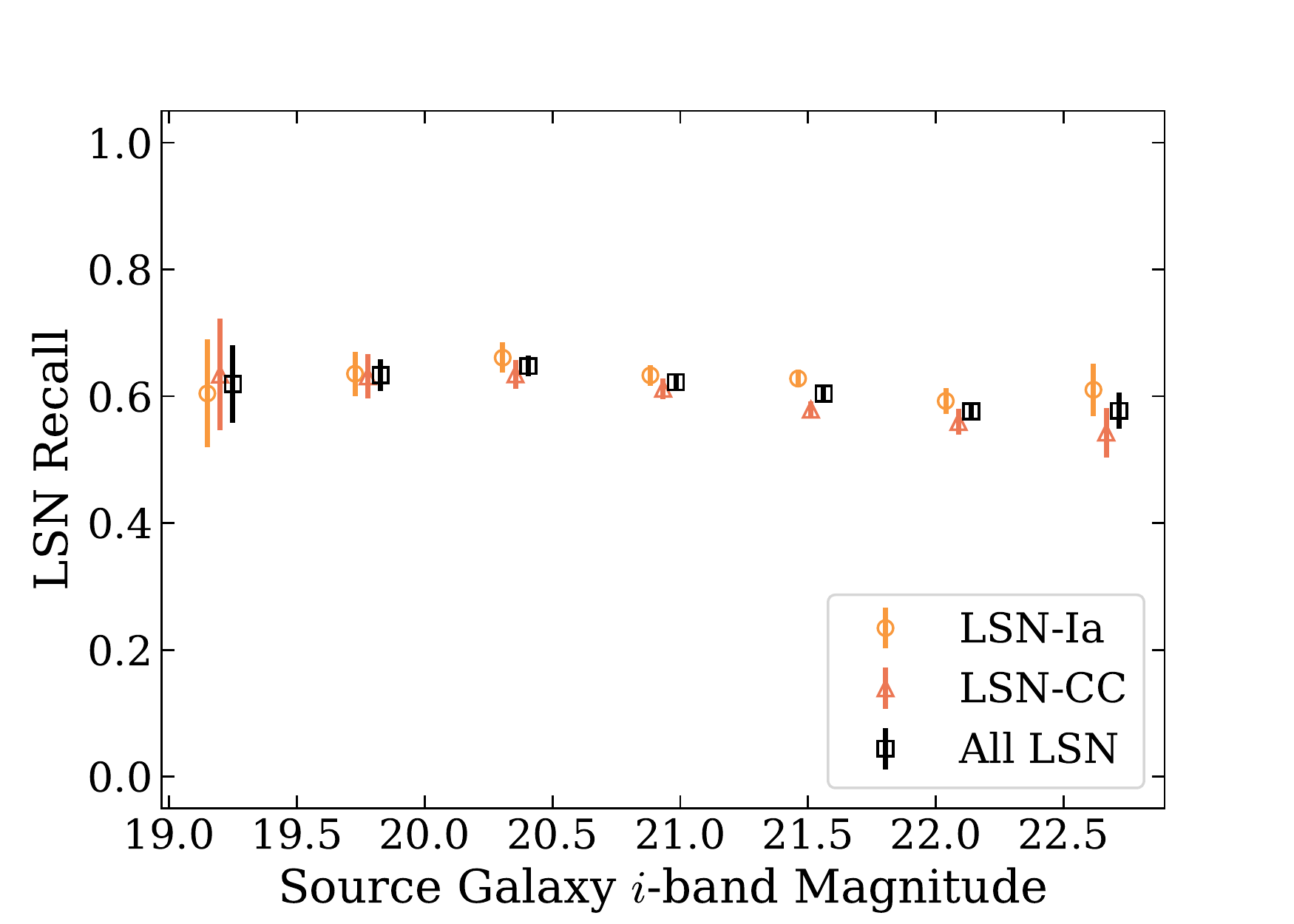}
    \includegraphics[width=0.49\textwidth]{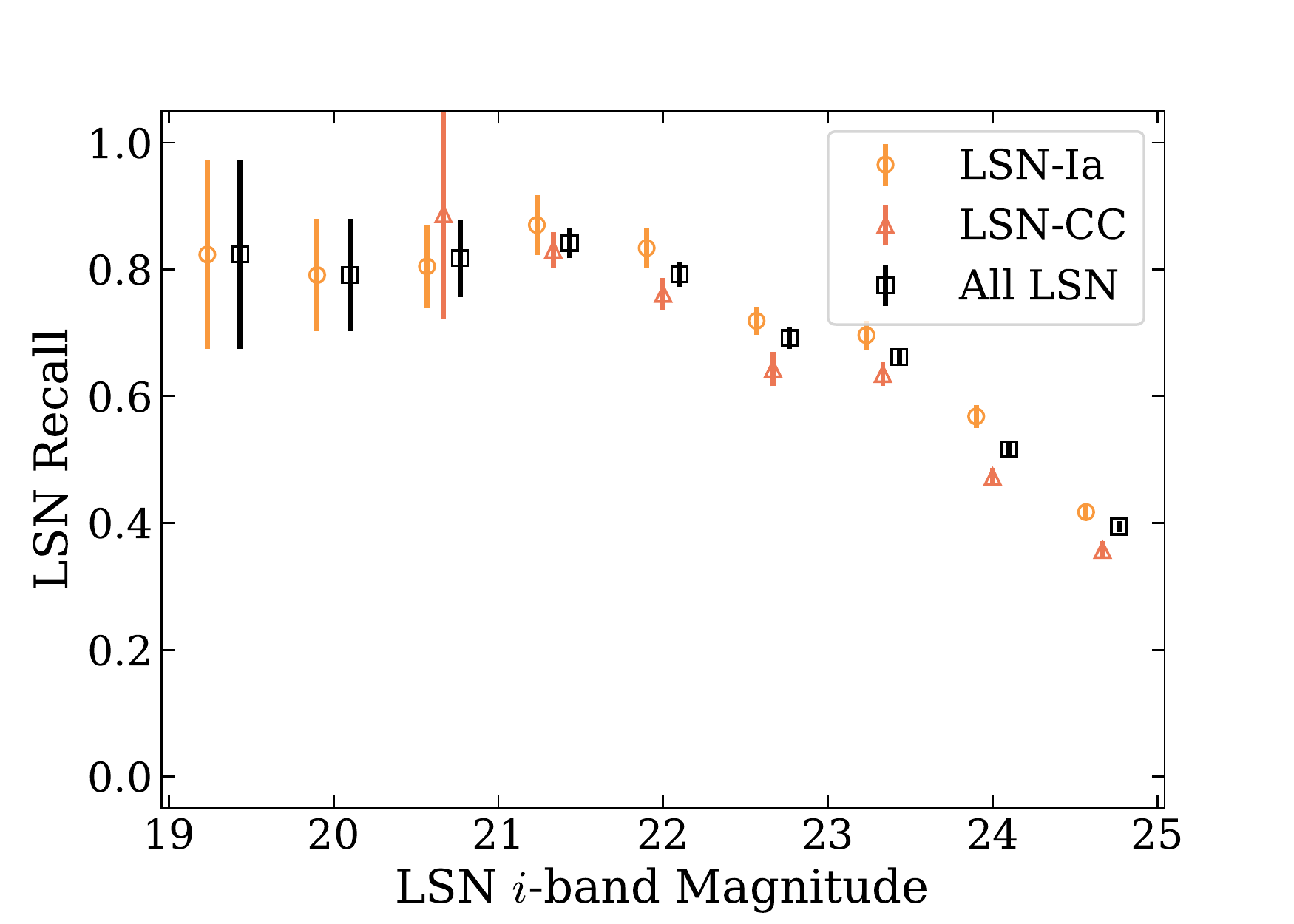}
    \caption{The physical selection function for simulated LSNe in our validation dataset. We measure the LSN recall (defined in Section \ref{sec:metrics}) as a function of Einstein radius, seeing, source galaxy unlensed magnitude, and LSN simulated unlensed magnitude. Error bars show a statistical uncertainty of one standard deviation.}
    \label{fig:selection}
\end{figure*}

We present the selection function for our ZipperNet and our selection criterion for the number of detections.
In this section, we analyze four central properties to understand the selection function of our approach: the Einstein radius, the seeing, the brightness of the source galaxy, and the brightness of the LSN.
The LSN recall is calculated as a function of these four properties in Figure \ref{fig:selection}.

We find that our approach has an easier time identifying larger Einstein radii than smaller Einstein radii.
Similarly, we find that our approach has an easier time identifying LSNe in good seeing conditions than in average or poor seeing conditions.
Both of these properties shed light the importance of the clarity of spatial features in the images.
Poor seeing or small Einstein radii are both situations in which image resolution is compromised and consequently spatial features become difficult or impossible to realize.
This trend in algorithm performance points to data quality characteristics as opposed to a selection bias introduced by our approach.
Lastly, we find that source galaxy brightness has little impact on the performance for the range applicable to this magnitude-limited analysis.
We observe a similar trend for LSNe brighter than 22~mag, but notice a reduction in performance for fainter LSNe.
This magnitude threshold is near the single-epoch limiting magnitude for the DES SN fields and is likely due to Malmquist bias, but there could be second-order selection effects in the detectability of the LSN images.

\label{app:selection}

\clearpage

\section{Candidate Metadata}
\label{app:table}
\begin{table*}
\hspace{-2.25cm}
    \begin{tabular}{|c|cccccccc|} \toprule
    \bf{No}. & \bf{Coadd Id.} & \bf{R.A. [deg.]} & \bf{Dec. [deg.]} & \bf{Mag$_i$} & \bf{Redshift} & \bf{Field} & \bf{Years Detected} & \bf{Inspection} \\ \midrule
1 & DES-691022126 & 53.898910 & -28.912293 & 21.73 & - & C2 & Y2 Y3 & 5 \\[-0.1cm]
2 & DES-701263907 & 40.969218 & -0.619054 & 17.31 & 0.030024 & S2 & Y5 & 5 \\[-0.1cm]
3 & DES-699919273 & 10.155917 & -44.437515 & 18.95 & 0.556 & E2 & Y5 & 5 \\[-0.1cm]
\midrule
4 & DES-690157493 & 53.394495 & -26.716658 & 21.82 & - & C1 & Y1 & 4 \\[-0.1cm]
5 & DES-690456076 & 55.010602 & -26.549543 & 20.79 & - & C1 & Y1 Y2 Y3 Y4 & 4 \\[-0.1cm]
6 & DES-690583502 & 55.066197 & -27.347839 & 21.44 & - & C1 & Y1 Y2 Y3 Y4 Y5 & 4 \\[-0.1cm]
7 & DES-690642061 & 54.464014 & -27.418446 & 21.62 & - & C1 & Y2 & 4 \\[-0.1cm]
8 & DES-690918939 & 54.410941 & -28.417641 & 20.65 & - & C2 & Y2 & 4 \\[-0.1cm]
9 & DES-691032289 & 54.420139 & -29.048729 & 21.80 & - & C2 & Y2 & 4 \\[-0.1cm]
10 & DES-691068769 & 53.476146 & -29.331470 & 20.89 & - & C2 & Y1 Y2 Y4 Y5 & 4 \\[-0.1cm]
11 & DES-691440047 & 55.134258 & -29.302642 & 20.72 & - & C2 & Y2 & 4 \\[-0.1cm]
12 & DES-691442767 & 55.214489 & -29.331276 & 21.79 & 0.139740 & C2 & Y4 & 4 \\[-0.1cm]
13 & DES-691524775 & 54.476519 & -29.498563 & 21.37 & - & C2 & Y2 & 4 \\[-0.1cm]
14 & DES-691664180 & 53.787326 & -29.869823 & 21.05 & - & C2 & Y5 & 4 \\[-0.1cm]
15 & DES-691702170 & 54.607503 & -29.829073 & 19.42 & - & C2 & Y4 & 4 \\[-0.1cm]
16 & DES-691896609 & 53.148074 & -27.405501 & 22.04 & 0.725 & C3 & Y1 Y2 Y4 & 4 \\[-0.1cm]
17 & DES-691902610 & 52.888864 & -27.157390 & 22.37 & - & C3 & Y2 & 4 \\[-0.1cm]
18 & DES-691947063 & 52.597080 & -27.675669 & 22.22 & - & C3 & Y1 & 4 \\[-0.1cm]
19 & DES-691968734 & 52.958718 & -27.916592 & 21.93 & 0.610106 & C3 & Y2 & 4 \\[-0.1cm]
20 & DES-692023723 & 52.061383 & -28.071403 & 22.20 & 0.949 & C3 & Y1 Y2 Y3 Y4 Y5 & 4 \\[-0.1cm]
21 & DES-692243027 & 53.581253 & -27.886280 & 21.18 & 0.739 & C3 & Y4 & 4 \\[-0.1cm]
22 & DES-692639734 & 53.269373 & -28.909563 & 21.01 & 0.471666 & C3 & Y2 Y4 & 4 \\[-0.1cm]
23 & DES-693331974 & 52.064812 & -28.509826 & 22.47 & - & C3 & Y2 Y5 & 4 \\[-0.1cm]
24 & DES-693351134 & 53.144038 & -28.581257 & 20.66 & 0.815070 & C3 & Y1 Y2 Y4 Y5 & 4 \\[-0.1cm]
25 & DES-695852037 & 34.616442 & -4.670152 & 21.25 & - & X1 & Y1 Y2 Y4 Y5 & 4 \\[-0.1cm]
26 & DES-696865317 & 35.170551 & -6.631484 & 20.56 & - & X2 & Y1 Y2 Y4 Y5 & 4 \\[-0.1cm]
27 & DES-697161182 & 35.442353 & -6.948239 & 22.25 & - & X2 & Y1 & 4 \\[-0.1cm]
28 & DES-697274399 & 36.589451 & -3.896260 & 20.39 & 0.435 & X3 & Y2 Y3 Y4 Y5 & 4 \\[-0.1cm]
29 & DES-697446876 & 36.990923 & -4.185003 & 22.47 & 0.463 & X3 & Y2 Y4 Y5 & 4 \\[-0.1cm]
30 & DES-697521552 & 36.249985 & -4.382867 & 22.23 & 0.798 & X3 & Y2 Y3 Y4 Y5 & 4 \\[-0.1cm]
31 & DES-698587357 & 7.138332 & -42.415813 & 21.02 & - & E1 & Y1 & 4 \\[-0.1cm]
32 & DES-698925976 & 9.245344 & -43.112557 & 20.18 & 0.318563 & E1 & Y2 Y4 & 4 \\[-0.1cm]
33 & DES-699088459 & 7.447012 & -43.647847 & 18.95 & - & E1 & Y2 & 4 \\[-0.1cm]
34 & DES-699127397 & 7.528665 & -43.465050 & 21.21 & 0.657900 & E1 & Y1 Y2 Y3 Y4 Y5 & 4 \\[-0.1cm]
35 & DES-699219206 & 9.869087 & -43.142798 & 20.47 & - & E2 & Y2 & 4 \\[-0.1cm]
36 & DES-699235372 & 9.611497 & -43.351731 & 20.46 & - & E2 & Y2 & 4 \\[-0.1cm]
37 & DES-699340227 & 10.241319 & -43.413401 & 20.32 & - & E2 & Y1 Y2 Y3 Y4 Y5 & 4 \\[-0.1cm]
38 & DES-699466457 & 9.455661 & -43.915105 & 22.19 & 0.469 & E2 & Y1 & 4 \\[-0.1cm]
39 & DES-699478563 & 8.862263 & -44.227511 & 20.67 & 0.751 & E2 & Y2 & 4 \\[-0.1cm]
40 & DES-699621639 & 8.923322 & -44.133374 & 19.12 & 0.235 & E2 & Y1 Y2 & 4 \\[-0.1cm]
41 & DES-699723043 & 10.869276 & -44.022876 & 21.71 & - & E2 & Y1 Y2 Y3 Y5 & 4 \\[-0.1cm]
42 & DES-699926736 & 8.351206 & -43.536464 & 21.14 & - & E1 & Y1 Y2 Y4 Y5 & 4 \\[-0.1cm]
43 & DES-700364825 & 42.449669 & 0.176652 & 20.46 & - & S1 & Y4 & 4 \\[-0.1cm]
44 & DES-700492744 & 42.157070 & -0.524332 & 19.30 & - & S1 & Y5 & 4 \\[-0.1cm]
45 & DES-700541568 & 41.136120 & -0.444152 & 21.63 & - & S2 & Y4 & 4 \\[-0.1cm]
46 & DES-700548040 & 41.369059 & -0.530169 & 21.29 & - & S2 & Y1 Y2 Y4 & 4 \\[-0.1cm]
47 & DES-700863020 & 41.521851 & -0.190829 & 20.52 & - & S2 & Y2 & 4 \\[-0.1cm]
48 & DES-700977591 & 41.022356 & -0.785606 & 18.75 & 0.287564 & S2 & Y1 Y2 Y5 & 4 \\[-0.1cm]
49 & DES-701328706 & 42.256074 & -1.098086 & 20.76 & - & S2 & Y1 & 4 \\[-0.1cm]
50 & DES-701662201 & 41.406713 & -1.916363 & 21.23 & - & S2 & Y3 & 4 \\[-0.1cm]\bottomrule
    \end{tabular}
    \caption{
    Properties of the systems detected by ZipperNet that received a score of 4 or 5 during the human visual inspection. 
    The ``Coadd Id.'' is from the DES Y3 GOLD Catalog. 
    The ``Years Detected'' indicate the years of DES data collection during which the candidate was selected by ZipperNet.
    The ``Redshift'' values are either photometric estimates from DES (shown to three significant digits) or spectroscopic measurements from OzDES \citep{ozdes} and refer to the candidate lensing galaxy.
    }
    \label{tab:candidates}
\end{table*}
This appendix lists properties of systems detected by ZipperNet and scored as a ``4'' or a ``5'' by human visual inspection.
\clearpage

\bibliographystyle{yahapj_twoauthor_arxiv_amp}
\bibliography{main}
\end{document}